\documentclass{article}

\usepackage[preprint,nonatbib]{nips_2018}

\usepackage[utf8]{inputenc} 
\usepackage{url}            
\usepackage{booktabs}       
\usepackage{amsfonts}       
\usepackage{amsmath}
\usepackage{amsthm}
\usepackage{nicefrac}       
\usepackage{microtype}      
\usepackage{enumitem}
\usepackage{bbm}
\usepackage{bm}
\usepackage{amssymb}
\usepackage{graphicx}
\usepackage[noend]{algorithmic}
\usepackage{algorithm}
\usepackage{wrapfig}
\usepackage{caption}
\usepackage{subcaption}
\usepackage{multirow}
\usepackage{booktabs}
\newcommand{\ra}[1]{\renewcommand{\arraystretch}{#1}}

\newtheorem{theorem}{Theorem}

\newcounter{lemmas}
\newtheorem{lemma}[lemmas]{Lemma}

\makeatletter
\renewenvironment{proof}[1][\proofname]{\par
  \vspace{-\topsep}
  \pushQED{\qed}%
  \normalfont
  \topsep0pt \partopsep0pt 
  \trivlist
  \item[\hskip\labelsep
        \itshape
    #1\@addpunct{.}]\ignorespaces
}{%
  \popQED\endtrivlist\@endpefalse
  \addvspace{0pt plus 0pt} 
}
\makeatother

\title{Inference Aided Reinforcement Learning for Incentive Mechanism Design in Crowdsourcing}

\author{
  Zehong Hu \\
  Nanyang Technological University\\
  \texttt{HUZE0004@e.ntu.edu.sg} \\
  \And
  Yitao Liang \\
  University of California, Los Angeles \\
  \texttt{yliang@cs.ucla.edu} \\
  \AND
  Yang Liu \\
  Harvard University \\
  \texttt{yangl@seas.harvard.edu} \\
  \And
  Jie Zhang \\
  Nanyang Technological University\\
  \texttt{ZhangJ@ntu.edu.sg} \\
}
\ifodd 1
\newcommand{\com}[1]{\textbf{\color{red}(COMMENT: #1)}} 
\newcommand{\clar}[1]{\textbf{\color{green}(NEED CLARIFICATION: #1)}}
\newcommand{\response}[1]{\textbf{\color{magenta}(RESPONSE: #1)}} 
\else

\newcommand{\com}[1]{}
\newcommand{\clar}[1]{}
\newcommand{\response}[1]{}
\fi

\begin{document}

\maketitle

\begin{abstract}
Incentive mechanisms for crowdsourcing are designed to incentivize financially self-interested workers to generate and report high-quality labels. Existing mechanisms are often developed as one-shot static solutions, assuming a certain level of knowledge about worker models (expertise levels, costs of exerting efforts, etc.). In this paper, we propose a novel inference aided reinforcement mechanism 
that learns to incentivize high-quality data sequentially and requires no such prior assumptions.
Specifically, we first design a Gibbs sampling augmented Bayesian inference algorithm to estimate workers' labeling strategies from the collected labels at each step. Then we propose a reinforcement incentive learning (RIL) method, building on top of the above estimates, to uncover how workers respond to different payments. RIL dynamically determines the payment without accessing any ground-truth labels. We theoretically prove that RIL is able to incentivize rational workers to provide high-quality labels. Empirical results show that our mechanism performs consistently well under both rational and non-fully rational (adaptive learning) worker models. Besides, the payments offered by RIL are more robust
and have lower variances compared to the existing one-shot mechanisms.
\end{abstract}

\section{Introduction}
The ability to quickly collect large-scale and high-quality labeled datasets is crucial for Machine Learning (ML). Among all proposed solutions, one of the most promising options is crowdsourcing \cite{Howe2006,slivkins2014online,difallah2015dynamics,simpson2015language}. Nonetheless, it has been noted that crowdsourced data often suffers from quality issue, due to its salient feature of no monitoring and no ground-truth verification of workers' contribution. This quality control challenge has been attempted by two relatively disconnected research communities. From the more ML side, quite a few inference techniques have been developed to infer true labels from crowdsourced and potentially noisy labels \cite{raykar2010learning,liu2012variational,zhou2014aggregating,zheng2017truth}. These solutions often work as one-shot, post-processing procedures facing a static set of workers, whose labeling accuracy is fixed and \emph{informative}.
Despite their empirical success, the aforementioned methods ignore the effects of \emph{incentives} when dealing with human inputs.
It has been observed both in theory and practice that, without appropriate incentive, selfish and rational workers tend to contribute low quality, uninformative, if not malicious data~\cite{sheng2008get,liu2017sequential}. Existing inference algorithms are very vulnerable to these cases - either much more redundant labels would be needed (low quality inputs), or the methods would simply fail to work (the case where inputs are uninformative and malicious). 

From the less ML side, the above quality control question has been studied in the context of \emph{incentive mechanism design}. In particular, a family of mechanisms, jointly referred as \emph{peer prediction}, have been proposed 
\cite{prelec2004bayesian,jurca2009mechanisms,witkowski2012peer,dasgupta2013crowdsourced}. Existing peer prediction mechanisms focus on achieving \emph{incentive compatibility} (IC), which is defined as that truthfully reporting private data, or reporting high quality data, maximizes workers' expected utilities. These mechanisms achieve IC via comparing the reports from the to-be-scored worker, against those from a randomly selected reference worker, to bypass the challenge of no ground-truth verification.
However, we note several undesirable properties of these methods.
Firstly, from learning's perspective, collected labels contain rich information about the ground-truth labels and workers' labeling accuracy.
Existing peer prediction mechanisms
often rely on reported data from a small subset of reference workers, which only represents a limited share of the overall collected information. In consequence, the mechanism designer dismisses the opportunity to leverage learning methods to generate a more credible and informative reference answer for the purpose of evaluation.
Secondly, existing peer prediction mechanisms often require a certain level of prior knowledge about workers' models, such as the cost of exerting efforts, and their labeling accuracy when exerting different levels of efforts. However, this prior knowledge is difficult to obtain under real environment. Thirdly, they often assume workers are all fully rational and always follow the utility-maximizing strategy. Rather, they may adapt their strategies in a dynamic manner.


In this paper, we propose an \emph{inference-aided reinforcement mechanism}, aiming to merge and extend the techniques from both inference and incentive design communities to address the caveats when they are employed alone, as discussed above.
The high level idea is as follows: we collect data in a sequential fashion. At each step, we assign workers a certain number of tasks and estimate the true labels and workers' strategies from their labels. Relying on the above estimates, a reinforcement learning (RL) algorithm is prosed to uncover how workers respond to different levels of offered payments.
The RL algorithm determines the payments for the workers based on the collected information up-to-date.
By doing so, our mechanism not only incentivizes (non-)rational workers to provide high-quality labels but also dynamically adjusts the payments according to workers' responses to maximize the data requester's cumulative utility. 
Applying standard RL solutions here is challenging, due to unobservable states (workers' labeling strategies) and reward (the aggregated label accuracy) which is further due to the lack of ground-truth labels. Leveraging standard inference methods seems to be a plausible solution at the first sight (for the purposes of estimating both the states and reward), but we observe that existing methods tend to over-estimate the aggregated label accuracy, which would mislead the superstructure RL algorithm.

We address the above challenges and make the following contributions: (1) We propose a Gibbs sampling augmented Bayesian inference algorithm, which estimates workers' labeling strategies and the aggregated label accuracy, as done in most existing inference algorithms, but significantly lowers the estimation bias of labeling accuracy. This lays a strong foundation for constructing correct reward signals, which are extremely important if one wants to leverage reinforcement learning techniques.
(2) A reinforcement incentive learning (RIL) algorithm is developed to maximize the data requester's cumulative utility by dynamically adjusting incentive levels according to workers' responses to payments. (3) We prove that our Bayesian inference algorithm and RIL algorithm are incentive compatible (IC) at each step and in the long run, respectively. (4) Experiments are conducted to test our mechanism, which shows that our mechanism performs consistently well under different worker models.
Meanwhile, compared with the state-of-the-art peer prediction solutions, our Bayesian inference aided mechanism can improve the robustness and lower the variances of payments.

\section{Related Work}
\label{section: related work}
Our work is inspired by the following three lines of literature:

\emph{Peer Prediction:} This line of work, addressing the incentive issues of eliciting high quality data without verification, starts roughly with the seminal ones \cite{prelec2004bayesian,gneiting2007strictly}. A series of follow-ups have relaxed various assumptions that have been made \cite{jurca2009mechanisms,witkowski2012peer,radanovic2013robust,dasgupta2013crowdsourced}. 

\emph{Inference method:} Recently, inference methods have been applied to crowdsourcing settings, aiming to uncover the true labels from multiple noisily copies. Notable successes include EM method \cite{dawid1979maximum,raykar2010learning,zhang2014spectral}, Variational Inference \cite{liu2012variational,chen2015statistical} and Minimax Entropy Inference~\cite{zhou2012learning,zhou2014aggregating}. Besides, Zheng \textit{et al.} \cite{zheng2017truth} provide a good survey for the existing ones.

\emph{Reinforcement Learning:} Over the past two decades, reinforcement learning (RL) algorithms have been proposed to iteratively improve the acting agent's learned policy \cite{Watkins92, Tesauro95, Sutton98, Gordon00, Szepesvari10}. More recently, with the help of advances in feature extraction and state representation, RL has made several breakthroughs in achieving human-level performance in challenging domains \cite{Mnih15,Liang16, Hasselt2016DeepRL, Silver17}. Meanwhile, many studies successfully deploy RL to address some societal problems \cite{Yu2013EmotionalMR,Leibo2017}. RL has also helped make progress in human-agent collaboration \cite{engel2005reinforcement, gasic2014gaussian,Sadhu2016ArgusSH,Wang2017}.


Our work differs from the above literature in the connection between incentive mechanisms and ML. There have been a very few recent studies that share a similar research taste with us.
For example, to improve the data requester's utility in crowdsourcing settings, Liu and Chen \cite{liu2017sequential} develop a multi-armed bandit algorithm to adjust the state-of-the-art peer prediction mechanism DG13~\cite{dasgupta2013crowdsourced} to a prior-free setting. Nonetheless, the results in above work require workers to follow a Nash Equilibrium at each step in a sequential setting, which is hard to achieve in practice. Instead of randomly choosing a reference worker as commonly done in peer prediction, Liu and Chen \cite{liu2017machine} propose to use supervised learning algorithms to generate the reference reports and derive the corresponding IC conditions. However, these reports need to be based on the contextual information of the tasks.
By contrast, in this paper, without assuming the contextual information about the tasks, we use Bayesian inference to learn workers' states and true labels, which leads to an unsupervised-learning solution.

\section{Problem Formulation}
\label{PF}
This paper considers the following data acquisition problem via crowdsourcing: at each discrete time step $t=1,2,...$, a data requester assigns $M$ tasks with binary answer space $\left\{-1,+1\right\}$ to $N \geq 3$ candidate workers to label. Workers receive payments for submitting a label for each task. We use $L^t_i(j)$ to denote the label worker $i$ generates for task $j$ at time $t$. For simplicity of computation, we reserve $L^t_i(j) = 0$ if  $j$ is not assigned to $i$. Furthermore, we use $\mathcal{L}$ and $\bm{L}$ to denote the set of ground-truth labels and  the set of all collected labels respectively.


The generated label $L^{t}_{i}(j)$ depends both on the latent ground-truth $\mathcal{L}(j)$ and worker $i$'s strategy, which is mainly determined by two factors: exerted effort level (high or low) and reporting strategy (truthful or deceitful).
Accommodating the notation commonly used in reinforcement learning, we also refer worker $i$'s strategy as his/her internal \emph{state}.
At any given time, workers at their will adopt an arbitrary combination of effort level and reporting strategy. Specifically, we define $\textsf{eft}^{t}_i\in[0,1]$ and $\textsf{rpt}^{t}_i\in[0,1]$ as worker $i$'s probability of exerting high efforts and reporting truthfully for task $j$, respectively.
Furthermore, we use $\mathbb{P}_{i,H}$ and $\mathbb{P}_{i,L}$ to denote worker $i$'s probability of observing the true label when exerting high and low efforts, respectively.
Correspondingly, we denote worker $i$'s cost of exerting high and low efforts by $c_{i,H}$ and $c_{i,L}$, respectively.
For the simplicity of analysis, we assume that $\mathbb{P}_{i,H}>\mathbb{P}_{i,L}=0.5$ and $c_{i,H}>c_{i,L}=0$. 
All the above parameters and workers' actions stay unknown to our mechanism.
In other words, we regard workers as black-boxes, which distinguishes our mechanism from the existing peer prediction mechanisms.

Worker $i$'s probability of being correct (PoBC) at time $t$ for any given task is given as
\begin{equation}
\label{PPP}
\begin{split}
\mathbb{P}^{t}_i  =& ~\textsf{rpt}^{t}_i \cdot\textsf{eft}^{t}_i \cdot \mathbb{P}_{i,H}+ (1-\textsf{rpt}^{t}_i)\cdot \textsf{eft}^{t}_i \cdot (1-\mathbb{P}_{i,H})+\\
&\quad\textsf{rpt}^{t}_i \cdot(1-\textsf{eft}^{t}_i) \cdot \mathbb{P}_{i,L}+(1-\textsf{rpt}^{t}_i) \cdot(1-\textsf{eft}^{t}_i)\cdot (1-\mathbb{P}_{i,L})
\end{split}
\end{equation}
Suppose we assign $m^{t}_i\leq M$ tasks to worker $i$ at step $t$. Then, a risk-neutral worker's utility satisfies:
\begin{equation}
\label{equation:u_of_worker}
u_i^t={\sum}_{j=1}^{M}P_i^t(j) - m^{t}_i \cdot c_{i,H}\cdot \textsf{eft}^{t}_i
\end{equation}
where $P^{t}_{i}$ denotes our payment to worker $i$ for task $j$ at time $t$ (see Section~\ref{section: methodology} for more details).

 
At the beginning of each step, the data requester and workers agree to a certain rule of payment, which is not changed until the next time step.
The workers are self-interested and may choose their strategies in labeling and reporting according to the expected utility he/she can get. 
After collecting the generated labels, the data requester infers the true labels $\tilde{L}^t(j)$ by running a certain inference algorithm.
The aggregated label accuracy $A^t$ and the data requester's utility $r_t$ are defined as follows:
\vspace{-3mm}
\begin{equation}
\label{equation:utility}
\begin{split}
A^t=\frac{1}{M}{\sum}_{j=1}^{M}1\left[\tilde{L}^{t}(j)=\mathcal{L}(j)\right]\;,\;
r_t= F(A^t) - \eta {\sum}_{i=1}^{N}{\sum}_{j=1}^{M}P^t_i(j)
\end{split}
\end{equation}
where $F(\cdot)$ is a non-decreasing monotonic function mapping accuracy to utility and $\eta>0$ is a tunable parameter balancing label quality and costs.



\section{Inference-Aided Reinforcement Mechanism for Crowdsourcing}
\label{section: methodology}
\begin{figure}[!htb]
\centering
\includegraphics[width=0.7\textwidth]{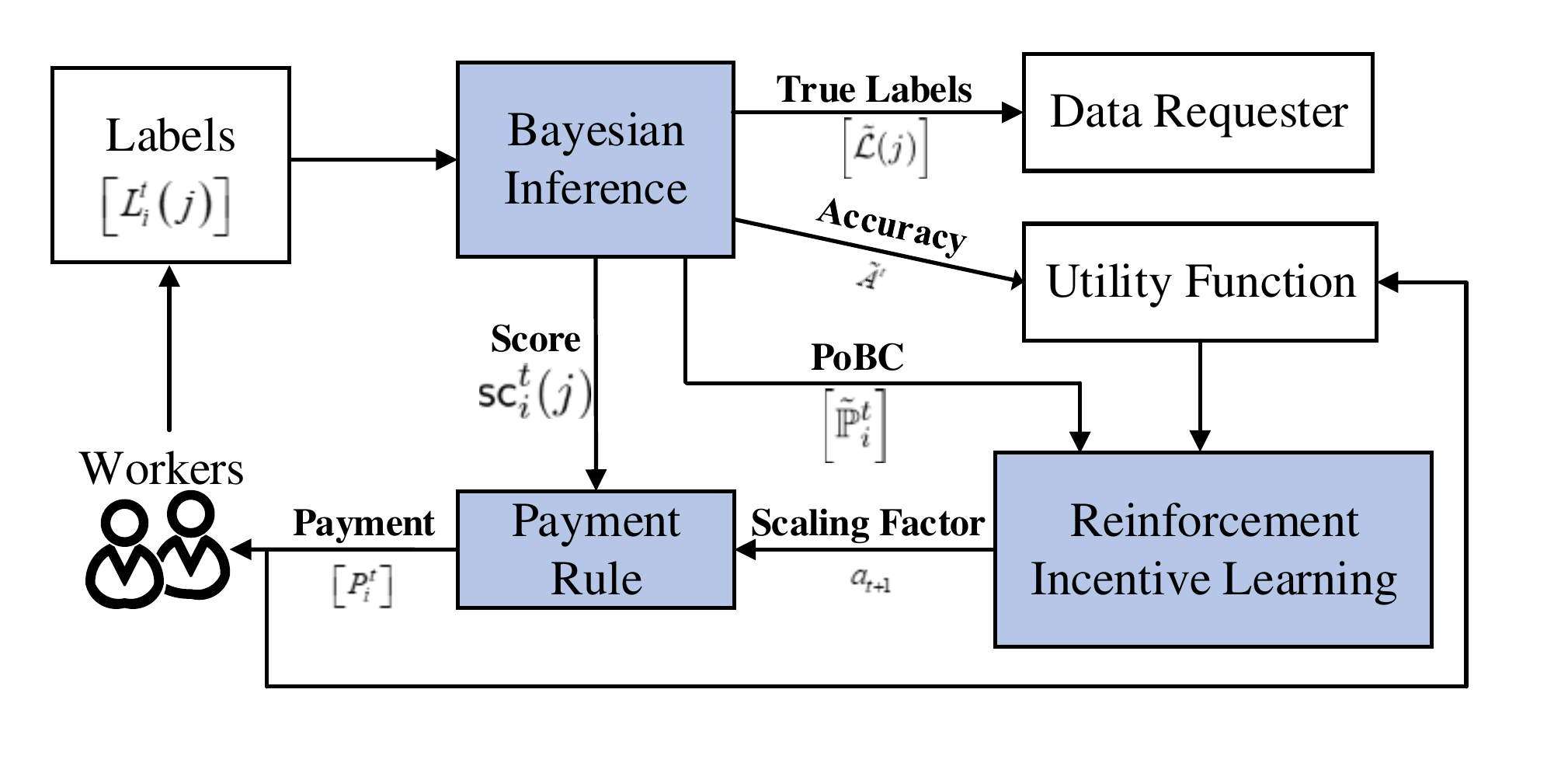}
\caption{\label{figure:layout} Overview of our incentive mechanism.}
\end{figure}
Our mechanism mainly consists of three components: the payment rule, Bayesian inference and reinforcement incentive learning (RIL); see Figure~\ref{figure:layout} for an overview, where estimated values are denoted with tildes.
The payment rule computes the payment to worker $i$ for his/her label on task $j$
\begin{equation}
P^t_i(j)=a_t \cdot [\textsf{sc}^{t}_i(j)-0.5]+b
\label{equation:payment}
\end{equation}
where $a_t \in \mathcal{A}$ denotes the scaling factor, determined by RIL at the beginning of every step $t$ and shared by all workers.
$\textsf{sc}^{t}_i(j)$ denotes worker $i$'s score on task $j$, which will be computed by the Bayesian inference algorithm.
$b\geq 0$ is a constant representing the fixed base payment. The Bayesian inference algorithm is also responsible for estimating the true labels, workers' PoBCs and the aggregated label accuracy at each time step, preparing the necessary inputs to RIL. Based on these estimates, RIL seeks to maximize the cumulative utility of the data requester by optimally balancing the utility (accuracy in labels) and the payments.

\subsection{Bayesian Inference}
\label{inference}
For the simplicity of notation, we omit the superscript $t$ in this subsection. The motivation for designing our own Bayesian inference algorithm is as follows. We ran several preliminary experiments using popular inference algorithms, for example, EM~\cite{dawid1979maximum,raykar2010learning,zhang2014spectral} and Variational Inference \cite{liu2012variational,chen2015statistical}). Our empirical studies reveal that those methods tend to heavily bias towards over-estimating the aggregated label accuracy when the quality of labels is low.\footnote{See Section 5.1 for detailed experiment results and analysis.}
This leads to biased estimation of the data requester's utility $r_t$ (as it cannot be observed directly), and this estimated utility is used as the reward signal in RIL, which will be detailed later.
Since the reward signal plays the core role in guiding the reinforcement learning process, the heavy bias will severely mislead our mechanism.

To reduce the estimation bias, we develop a Bayesian inference algorithm by introducing soft Dirichlet priors to both the distribution of true labels $\bm{\tau}=[\tau_{-1},\tau_{+1}]\sim \textrm{Dir}(\beta_{-1},\beta_{+1})$, where $\tau_{-1}$ and $\tau_{+1}$ denote that of label $-1$ and $+1$, respectively, and workers' PoBCs  $[\mathbb{P}_{i}, 1-\mathbb{P}_i]\sim \textrm{Dir}(\alpha_{1},\alpha_{2})$. Then, we derive the conditional distribution of true labels given collected labels as (see Appendix A)
$\mathbb{P}(\mathcal{L}|\bm{L})=\mathbb{P}(\bm{L},\mathcal{L})/\mathbb{P}(\bm{L})\propto B(\hat{\bm{\beta}}) \cdot {\prod}_{i=1}^{N}B(\hat{\bm{\alpha}}_{i}),
$ 
%
where $B(x,y)=(x-1)!(y-1)!/(x+y-1)!$ denotes the beta function, $\hat{\bm{\alpha}}=[\hat{\alpha}_1,\hat{\alpha}_2]$, $\hat{\bm{\beta}}=[\hat{\beta}_{-1},\hat{\beta}_{+1}]$, $\hat{\alpha}_{i1}={\sum}_{j=1}^{M}{\sum}_{k\in\{-1,+1\}}\delta_{ijk}\xi_{jk}+2\alpha_{1}-1$, $\hat{\alpha}_{i2}={\sum}_{j=1}^{M}{\sum}_{k\in\{-1,+1\}}\delta_{ij(-k)}\xi_{jk}+2\alpha_{2}-1$ and $\hat{\beta}_k={\sum}_{j=1}^{M}\xi_{jk}+2\beta_{k}-1$, where $\delta_{ijk}=\mathbbm{1}(L_i(j)=k)$ and $\xi_{jk}= \mathbbm{1}(\mathcal{L}(j)=k)$. 

  \begin{wrapfigure}{R}{0.6\textwidth}
  \centering
    \begin{minipage}{0.6\textwidth}
\begin{algorithm}[H]
  \caption{Gibbs sampling for crowdsourcing}
  \label{GSC}
  \small
\begin{algorithmic}[1]
  \vspace{0.5mm}
  \STATE {\bfseries Input:} the collected labels $\bm{L}$, the number of samples $W$
  \STATE {\bfseries Output:} the sample sequence $\mathcal{S}$
  \vspace{0.5mm}
  \STATE $\mathcal{S}\leftarrow\emptyset$, Initialize $\mathcal{L}$ with the uniform distribution
  \FOR{$s=1$ {\bfseries to} $W$}
  \FOR{$j=1$ {\bfseries to} $M$}
  \STATE $\mathcal{L}(j) \leftarrow 1$ and compute $x_1= B(\hat{\bm{\beta}})\prod_{i=1}^{N}B(\hat{\bm{\alpha}}_{i})$
  \STATE $\mathcal{L}(j) \leftarrow 2$ and compute $x_2= B(\hat{\bm{\beta}})\prod_{i=1}^{N}B(\hat{\bm{\alpha}}_{i})$
  \STATE $\mathcal{L} \leftarrow$ Sample $\{1,2\}$ with $P(1)=x_1/(x_1+x_2)$
  \ENDFOR
  \STATE Append $\tilde{\mathcal{L}}$ to the sample sequence $\mathcal{S}$
  \ENDFOR
\end{algorithmic}
\end{algorithm}
    \end{minipage}
  \end{wrapfigure}
Note that it is generally hard to derive an explicit formula for the posterior distribution of a specific task $j$'s ground-truth from the conditional distribution $\mathbb{P}(\mathcal{L}|\bm{L})$. We thus resort to Gibbs sampling for the inference.
More specifically, according to Bayes' theorem, we know that the conditional distribution of task $j$'s ground-truth $\mathcal{L}(j)$ satisfies
$\mathbb{P}[\mathcal{L}(j)|\bm{L}, \mathcal{L}(-j)]\propto \mathbb{P}(\mathcal{L}|\bm{L})$, where $-j$ denotes all tasks excluding $j$.
Leveraging this, we generate samples of the true label vector $\mathcal{L}$ following Algorithm~\ref{GSC}.
At each step of the sampling procedure (lines 6-7), Algorithm~\ref{GSC} first computes $\mathbb{P}[\mathcal{L}(j)|\bm{L}, \mathcal{L}(-j)]$ and then generates a new sample of $\mathcal{L}(j)$ to replace the old one in $\tilde{\mathcal{L}}$.
After traversing through all tasks, Algorithm~\ref{GSC} generates a new sample of the true label vector $\mathcal{L}$.
Repeating this process for $W$ times, we get $W$ samples, which is recorded in $\mathcal{S}$.
Here, we write the $s$-th sample as $\tilde{\mathcal{L}}^{(s)}$.
Since Gibbs sampling requires a burn-in process, we discard the first $W_0$ samples and calcualte worker $i$'s score on task $j$ and PoBC as
\begin{equation}
\label{equaton:score}
\textsf{sc}^{t}_i(j) = {\sum}_{s=W_0}^{W}\frac{\mathbbm{1}\left[\tilde{\mathcal{L}}^{(s)}(j)=L_{i}(j)\right]}{W-W_0}, \tilde{\mathbb{P}}_{i}=\frac{\sum\limits_{s=W_0}^{W}\bigl[2\alpha_{1}-1+\sum\limits_{j=1}^{M}\mathbbm{1}(\tilde{\mathcal{L}}^{(s)}(j)=L_{i}(j))\bigr]}{(W-W_0)\cdot(2\alpha_{1}+2\alpha_{2}-2+m_i)}.
\end{equation}
Similarly, we can obtain the estimates of the true label distribution $\bm{\tau}$ and then derive the log-ratio of task $j$, $\sigma_j=\log(\mathbb{P}[\mathcal{L}(j)=-1]/\mathbb{P}[\mathcal{L}(j)=+1])$.
Furthermore, we decide the true label estimate $\tilde{\mathcal{L}}(j)$ as $-1$ if $\tilde{\sigma}_j>0$ and as $+1$ if $\tilde{\sigma}_j<0$.
Correspondingly, the label accuracy $A$ is estimated as
\begin{equation}
\label{vot}
\begin{split}
\tilde{A}=\mathbb{E}\left(A \right) = M^{-1}{\sum}_{j=1}^{M}e^{|\tilde{\sigma}_j|}\left(1+e^{|\tilde{\sigma}_j|}\right)^{-1}.
\end{split}
\end{equation}

In our Bayesian inference algorithm, workers' scores, PoBCs and the true label distribution are all estimated by comparing the true label samples with the collected labels. 
Thus, t
To prove the convergence of our algorithm, we need to bound the ratio of wrong samples.
We introduce $n$ and $m$ to denote the number of tasks of which the true label sample in Eqn. (\ref{equaton:score}) is correct ($\tilde{\mathcal{L}}^{(s)}(j)=\mathcal{L}(j)$) and wrong ($\tilde{\mathcal{L}}^{(s)}(j)\neq \mathcal{L}(j)$) in the $s$-th sample, respectively.
Formally, we have: 
\begin{lemma}
\label{ConvBound}
Let $\bar{\mathbb{P}}=1-\mathbb{P}$, $\hat{\mathbb{P}}=\max \{\mathbb{P}, \bar{\mathbb{P}}\}$ and $\mathbb{P}_0=\tau_{-1}$. When $M\gg 1$,
\begin{equation}
\label{equation:CB}
\mathbb{E}[m/M]\lesssim (1+e^{\delta})^{-1}(\varepsilon+e^{\delta})(1+\varepsilon)^{M-1}\;,\;
\mathbb{E}[m/M]^2\lesssim (1+e^{\delta})^{-1}(\varepsilon^2+e^{\delta})(1+\varepsilon)^{M-2}
\end{equation}
where $\varepsilon^{-1}=\prod_{i=0}^{N}(2\hat{\mathbb{P}}_i)^{2}$, $\delta=O[\Delta\cdot \log(M)]$ and 
$\Delta={\sum}_{i=1}^N[1(\mathbb{P}_i<0.5)-1(\mathbb{P}_i>0.5)]$.
\end{lemma}
The proof is in Appendix B. Our main idea is to introduce a set of counts for the collected labels and then calculate $\mathbb{E}[m/M]$ and $\mathbb{E}[m/M]^2$ based on the distribution of these counts. Using Lemma~\ref{ConvBound}, the convergence of our Bayesian inference algorithm states as follows:


\begin{theorem}[Convergence]
\label{label:convergence}
When $M\gg 1$ and ${\prod}_{i=0}^{N}(2\hat{\mathbb{P}}_{i})^{2} \geq M$, if most of workers report truthfully (i.e. $\Delta<0$), with probability at least $1-\delta\in (0,1)$, $|\tilde{\mathbb{P}}_i-\mathbb{P}_i|\leq O(1/\sqrt{\delta M})$ holds for any worker $i$'s PoBC estimate $\tilde{\mathbb{P}}_i$ as well as the true label distribution estimate ($\tilde{\tau}_{-1}=\tilde{\mathbb{P}}_0$).
\end{theorem}
The convergence of $\tilde{\mathbb{P}}_i$ and $\tilde{\bm{\tau}}$ can naturally lead to the convergence of $\tilde{\sigma}_j$ and $\tilde{A}$ because the latter estimates are fully computed based on the former ones. 
All these convergence guarantees enable us to use the estimates computed by Bayesian inference to construct the state and reward signal in our reinforcement learning algorithm RIL.

\subsection{Reinforcement Incentive Learning}
\label{RL}
In this subsection, we formally introduce our reinforcement incentive learning (RIL) algorithm, which adjusts the scaling factor $a_t$ to maximize the data requesters' utility accumulated in the long run.
To fully understand the technical background, readers are expected to be familiar with Q-value and function approximation.
For readers with limited knowledge, we kindly refer them to Appendix D, where we provide background on these concepts.
With transformation, our problem can be perfectly modeled as a Markov Decision Process.
To be more specific, our mechanism is the agent and it interacts with workers (i.e. the environment); scaling factor $a_t$ is the action; the utility of the data requester $r_t$ defined in Eqn. (\ref{equation:utility}) is the reward.
Workers' reporting strategies are the state.
After receiving payments, workers may change their strategies to, for example, increase their utilities at the next step.
How workers change their strategies forms the state transition kernel.

On the other hand, the reward $r_t$ defined in Eqn. (\ref{equation:utility}) cannot be directly used because the true accuracy $A^t$ cannot be observed.
Thus, we use the estimated accuracy $\tilde{A}$ calculated by Eqn. (\ref{vot}) instead to approximate $r_t$ as in Eqn. (\ref{equation:approx_reward}).
Furthermore, to achieve better generalization across different states, it is a common approach to learn a feature-based state representation $\phi(s)$ \cite{Mnih15, Liang16}. Recall that the data requester's implicit utility at time $t$ only depends on the aggregated PoBC averaged across all workers. Such observation already points out to a  representation design with good generalization, namely 
$\phi(s_t) = {\sum}_{i=1}^N \mathbb{P}^t_i/N$.
Further recall that, when deciding the current scaling factor $a_t$, the data requester does not observe the latest workers' PoBCs and thus cannot directly estimate the current $\phi(s_t)$. Due to this one-step delay, we have to build our state representation using the previous observation. Since most workers would only change their internal states after receiving a new incentive, there exists some imperfect mapping function $\phi(s_{t}) \approx f(\phi(s_{t-1}),a_{t-1})$. 
Utilizing this implicit function, we introduce the augmented state representation in RIL as $\hat{s}_t$ in Eqn. (\ref{equation:approx_reward}).
\begin{equation}
\label{equation:approx_reward}
r_t\approx F(\tilde{A}^t) - \eta {\sum}_{i=1}^{N}P^t_i\;\;,\;\;\hat{s}_t = \langle \phi(s_{t-1}), a_{t-1} \rangle.
\end{equation}

Since neither $r_t$ nor $s_t$ can be perfectly inferred, it would not be a surprise to observe some noise that cannot be directly learned in our Q-function. 
For most crowdsourcing problems the number of tasks $M$ is large, so we can leverage the central limit theorem to justify our modeling of the noise using a Gaussian process.
To be more specific, we calculate the temporal difference (TD) error as 
\begin{equation}
r_t \approx Q^\pi(\hat{s}_t, a_t) - \gamma \mathbb{E}_{\pi}Q^{\pi}(\hat{s}_{t+1},a_{t+1}) + \epsilon_t 
\end{equation}

\vspace{-2mm}
  \begin{wrapfigure}{R}{0.6\textwidth}
  \centering
    \begin{minipage}{0.6\textwidth}
      {\centering
      \begin{algorithm}[H]
         \caption{Reinforcement Incentive Learning (RIL)}
         \label{RAC}

      \small
      \begin{algorithmic}[1]
         \FOR{each episode}
         \FOR{each step in the episode}
         \STATE Decide the scaling factor as ($\epsilon$-greedy method)
                  \vspace{-2mm}
                  $$\ \ a_t=\left\{
                  \begin{array}{ll}
                      \arg\max_{a\in\mathcal{A}}Q(\hat{s}_t,a) & \mathrm{Probability\ } 1-\epsilon\\
                      \mathrm{Random\ } a\in\mathcal{A} & \mathrm{Probability\ } \epsilon
                  \end{array}						
                   \right.$$  
                   \vspace{-3mm} 
         \STATE Assign tasks and collect labels from the workers
         \STATE Run Bayesian inference to get $\hat{s}_{t+1}$ and $r_t$
         \STATE Use $(\hat{s}_t, a_t, r_t)$ to update $\bm{K}$, $\bm{H}$ and $\bm{r}$ in Eqn. (\ref{equation:update})
         \ENDFOR
         \ENDFOR
      \end{algorithmic}
      \end{algorithm}}
    \end{minipage}
  \end{wrapfigure}
where the noise $\epsilon_t $ follows a Gaussian process, and $\pi=\mathbb{P}(a|\hat{s})$ denotes the current policy.
By doing so, we gain two benefits. First, the approximation greatly simplifies the derivation of the update equation for the Q-function. Secondly, as shown in our empirical results later, this kind of approximation is robust against different worker models.
Besides, following \cite{gasic2014gaussian} we approximate Q-function as
$Q^{\pi}(\hat{s}_{t+1},a_{t+1})\approx\mathbb{E}_{\pi}Q^{\pi}(\hat{s}_{t+1},a_{t+1})+\epsilon_{\pi}$, 
where $\epsilon_{\pi}$ also follows a Gaussian process.

Under the Gaussian process approximation, all the observed rewards and the corresponding $Q$ values up to the current step $t$ form a system of equations, and it can be written as
$\bm{r}=\bm{H}\bm{Q}+\bm{N}$, where $\bm{r}$, $\bm{Q}$ and $\bm{N}$ denote the collection of rewards, $Q$ values, and residuals. Following Gaussian process's assumption for residuals, $\bm{N}\sim \mathcal{N}(\bm{0},\bm{\sigma}^2)$, where $\bm{\sigma}^2=\textrm{diag}(\sigma^2,\ldots,\sigma^2)$.
The matrix $\bm{H}$ satisfies $\bm{H}(k,k)=1$ and $\bm{H}(k,k+1)=-\gamma$ for $k=1,\ldots, t$.
Then, by using the online Gaussian process regression algorithm~\cite{engel2005reinforcement}, we effectively learn the Q-function as
\begin{equation}
\label{equation:update}
Q(\hat{s},a) = \bm{k}(\hat{s},a) ^{\mathrm{T}}(\bm{K} +\bm{\sigma}^2)^{-1}\bm{H}^{-1}\bm{r}
\end{equation}
where $\bm{k}(\hat{s},a)=[k((\hat{s},a), (\hat{s}_1,a_1)),\ldots, k((\hat{s},a), (s_t,a_t))]^{\mathrm{T}}$ and $\bm{K}=[\bm{k}(\hat{s}_1,a_1),\ldots,\bm{k}(\hat{s}_t,a_t)]$. Here, we use $k(\cdot, \cdot)$ to denote the Gaussian kernel.
Finally, we employ the classic $\epsilon$-greedy method to decide $a_t$ based on the learned Q-function.
To summarize, we provide a formal description about RIL in Algorithm~\ref{RAC}. Note that, when updating $\bm{K}$, $\bm{H}$ and $\bm{r}$ in Line 6, we employ the sparse approximation proposed in \cite{gasic2014gaussian} to discard some data so that the size of these matrices does not increase infinitely.

\section{Theoretical Analysis on Incentive Compatibility}
\label{analysis}
In this section, we prove the incentive compatibility of our Bayesian inference and reinforcement learning algorithms. Our main results are as follows:
\begin{theorem}[One Step IC]
\label{OSEqulibrium}
At any time step $t$, when 
$M\gg 1,~{\prod}_{i=1}^{N}(2\mathbb{P}_{i,H})^{2} \geq M,~a_t>\max_{i}c_{i,H}/(\mathbb{P}_{i,H}-0.5)$, 
reporting truthfully and exerting high efforts is the utility-maximizing strategy for any worker $i$ at equilibrium (if other workers all follow this strategy).
\begin{proof}
In Appendix~E, we prove that when $a_t>c_{i,H}/(\mathbb{P}_{i,H}-0.5)$, if $\tilde{\mathbb{P}}^t_i\approx \mathbb{P}^t_i$, any worker $i$'s utility-maximizing strategy would be reporting truthfully and exerting high efforts.
Since Theorem 1 has provided the convergence guarantee, we can conclude Theorem 2.
\end{proof}
\end{theorem}
\begin{theorem}[Long Term IC]
\label{RMNE}
Suppose the conditions in Theorem~\ref{OSEqulibrium} are satisfied and the learned $Q$-function approaches the real $Q^{\pi}(\hat{s},a)$. When the following equation holds for $i=1,\ldots, N$,
\begin{equation}
\label{Condition}
\eta M\sum_{x\neq i} \mathbb{P}_{x,H} \cdot G_{\mathcal{A}}> \frac{F(1)-F(1-\psi_{i})}{1-\gamma}
\;,\; \psi_i =\left(\frac{\tau_{-1}}{\tau_{+1}}+\frac{\tau_{+1}}{\tau_{-1}}\right)\prod_{x\neq i}\sqrt{4\mathbb{P}_{x,H}(1-\mathbb{P}_{x,H})}
\end{equation}
always reporting truthfully and exerting high efforts is the utility-maximizing strategy for any worker $i$ in the long term if other workers all follow this strategy. Here, $G_{\mathcal{A}}=\min_{a,b\in\mathcal{A}, a\neq b}|a-b|$ denotes the minimal gap between two available values of the scaling factor.
\end{theorem}
In order to induce RIL to change actions, worker $i$ must let RIL learn a wrong $Q$-function. Thus, our main idea of proof is to derive the upper bounds of the effects of worker $i$'s reports on the $Q$-function.
Besides, Theorem 3 points that, to design robust reinforcement learning algorithms against the manipulation of strategical agents, we should leave a certain level of gaps between actions. This observation may be of independent interests to reinforcement learning researchers.

\section{Empirical Experiments} 


In this section, we empirically investigate the competitiveness of our solution. To be more specific, we first show our proposed Bayesian inference algorithm can produce more accurate estimates about the aggregated label accuracy when compared with the existing inference algorithms. Then, we demonstrate that, aided by Bayesian inference, our RIL algorithm consistently manages to learn a good incentive policy under various worker models. Lastly, we show as a bonus benefit of our mechanism that, leveraging Bayesian inference to fully exploit the information contained in the collected labels leads to more robust and lower-variance payments at each step.

\subsection{Empirical Analysis on Bayesian Inference}

The aggregated label accuracy estimated from our Bayesian inference algorithm serves as a major component of the state representation and reward function to RIL, and thus critically affects the performance of our mechanism.
Given so, we choose to first investigate the bias of our Bayesian inference algorithm.
In Figure~\ref{inference bias}, we compare our Bayesian inference algorithm with two popular inference algorithms in crowdsourcing,
that is, the EM estimator \cite{raykar2010learning} and the variational inference estimator \cite{liu2012variational}.
Here, we employ the famous RTE dataset, where workers need to check whether a hypothesis sentence can be inferred from the provided sentence~\cite{snow2008cheap}.
In order to simulate strategic behaviors of workers, we mix these data with random noise by replacing a part of real-world labels with uniformly generated ones (low quality labels).

From the figure, we conclude that compared with EM and variational inference, our Bayesian inference algorithm can significantly lower the bias of the estimates of the aggregated label accuracy. In fact, we cannot use the estimates from the EM and variational inference as alternatives for the reward signal because the biases of their estimates even reach $0.45$ while the range of the label accuracy is only between $[0.5, 1.0]$. This set of experiments justifies our motivation to develop our own inference algorithm and reinforces our claim that our inference algorithm could provide fundamentals for the further development of potential learning algorithms for crowdsourcing.



\begingroup
\vspace{3mm}
\begin{figure}[t]
    \centering
    \begin{subfigure}[t]{0.32\textwidth}
        \centering
        \includegraphics[width=\textwidth]{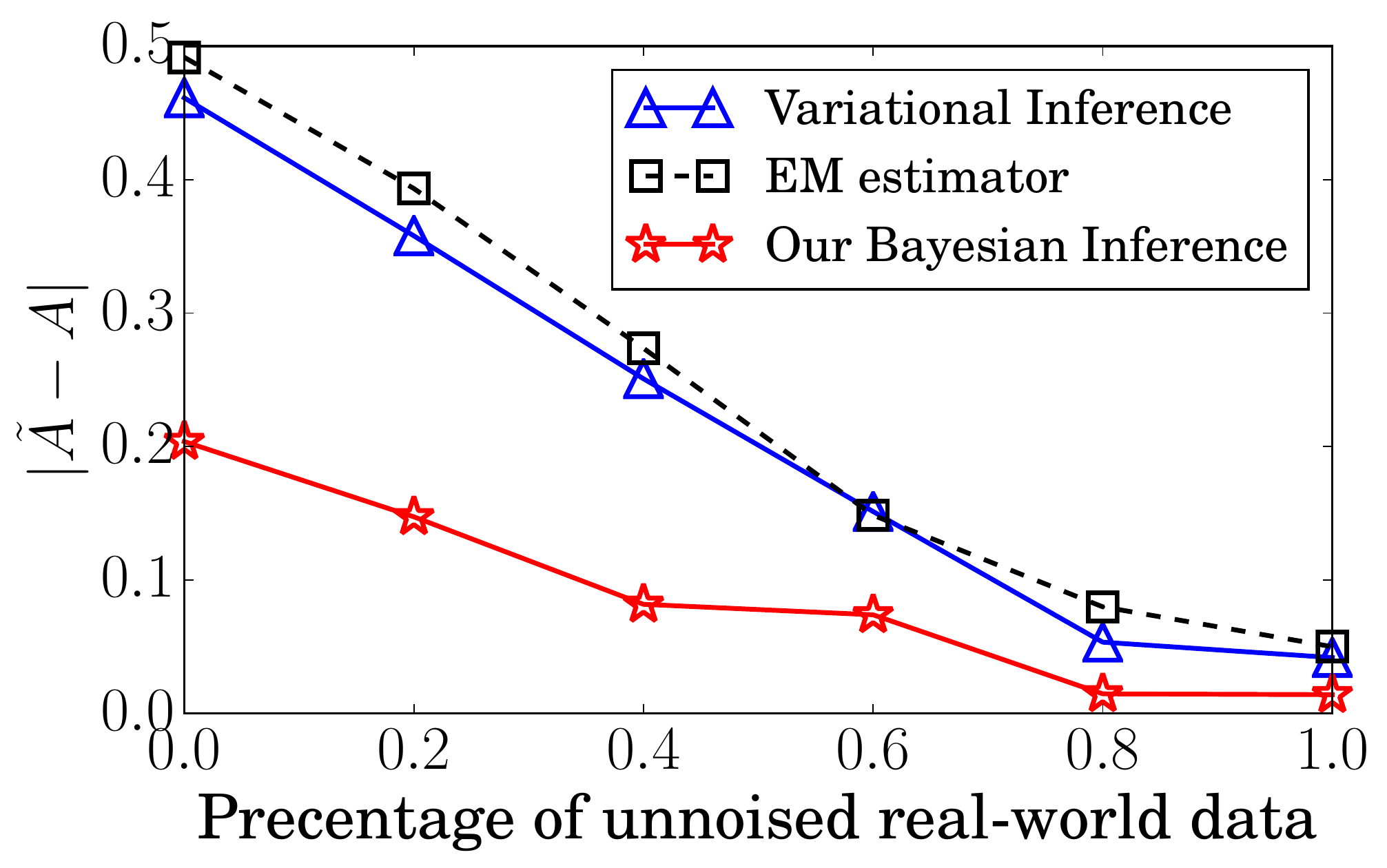}
	\caption{\label{inference bias}}
    \end{subfigure}%
    ~
    \begin{subfigure}[t]{0.32\textwidth}
        \centering
       \includegraphics[width=\textwidth]{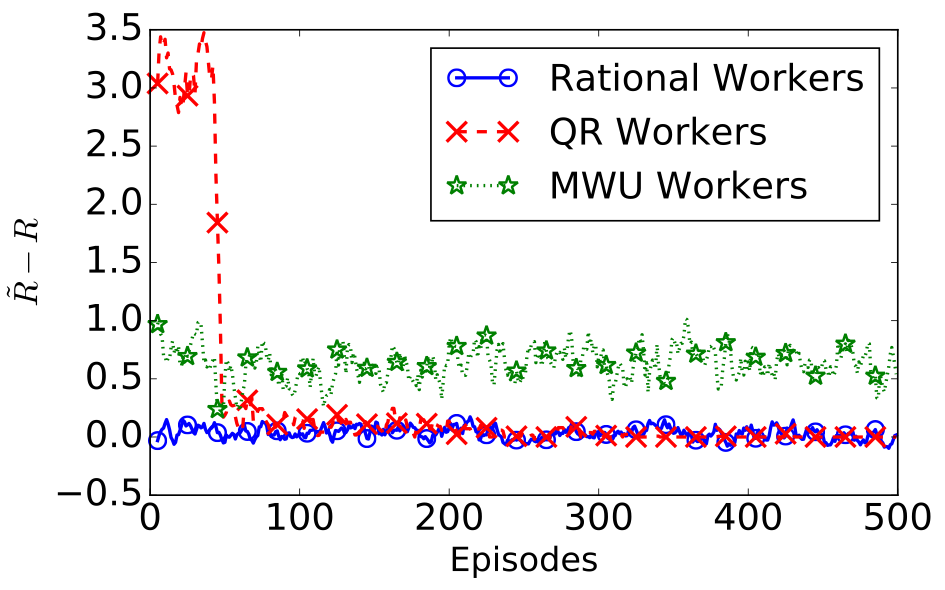}
        \caption{\label{figure:rewardError}}

    \end{subfigure}
        ~
    \begin{subfigure}[t]{0.32\textwidth}
        \centering
        \includegraphics[width=\textwidth]{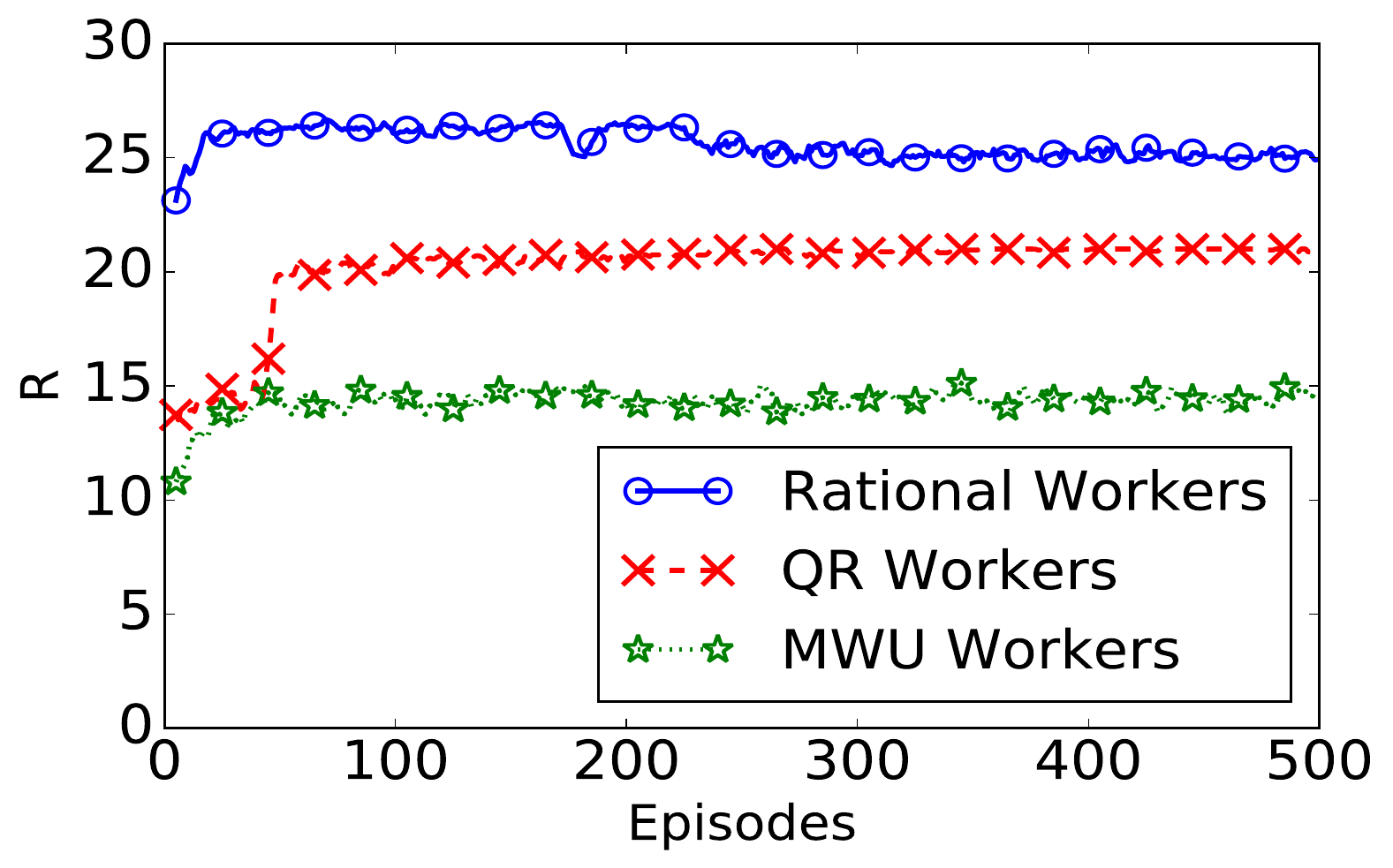}
                \caption{\label{figure:learningCurve}}
    \end{subfigure}
\caption{Empirical analysis on Bayesian Inference (a) and RIL (b-c). To be more specific, (a) compares the inference bias (i.e. the difference from the inferred label accuracy to the real one)  of our Bayesian inference algorithm with that of EM and variational inference, averaged over 100 runs. (b) draws the gap between the estimation of the data requester’s cumulative utility and the real one, smoothed over 5 episodes. (c) shows the learning curve of our mechanism, smoothed over 5 episodes.}
\vspace{-5mm}
\end{figure}

      \begin{table}[t]
      \ra{1.05}
          \caption{Performance comparison under three worker models. Data requester's cumulative utility normalized over the number of tasks. Standard deviation reported in parenthesis.}
          \label{table:performance}
         	\vspace{3mm}
          \centering
          {\footnotesize
          \begin{sc}
          \begin{tabular}{ @{}l c c c@{} }
          \toprule
          Method & Rational & QR & MWU \\ \midrule
          Fixed Optimal & 27.584 (.253) & 21.004 (.012) & 11.723 (.514) \\
          Heuristic Optimal & 27.643 (.174) & 21.006 (.001) & 12.304 (.515)\\
          Adaptive Optimal & 27.835 (.209) & 21.314 (.011) & 17.511 (.427) \\
          \midrule
          RIL & 27.184 (.336) & 21.016 (.018) & 15.726 (.416)\\
		\bottomrule
		\end{tabular}
          \end{sc}
          }
          \vspace{-4mm}
   \end{table}
\endgroup

\subsection{Empirical Analysis on RIL}
We move on to investigate whether RIL consistently learns a good policy, which maximizes the data requester's cumulative utility $R=\sum_t r_t$.  For all the experiments in this subsection, we set $M=100$, $N=10$, $\mathbb{P}_H=0.9$, $b=0$, $c_H=0.02$, the 
set of the scaling factor $\mathcal{A}=\{0.1,1.0,5.0,10\}$, the exploration rate $\epsilon = 0.2 $ for RIL and  $F(A)=A^{10}$, $\eta=0.001$ for the utility function (Eqn. (\ref{equation:utility})) and the number of time steps for an episode as $28$. We report the averaged results over 5 runs to reduce the effect of outliers. To demonstrate our algorithm's general applicability, we test it under three different worker models, each representing a popular family of human behavioral model.
We provide a simple description of them as follows, whereas the detailed version is deferred to Appendix H. (i) \emph{Rational} workers alway take the utility-maximizing strategies. (ii) \emph{QR} workers \cite{mckelvey1995quantal} follow strategies corresponding to an utility-dependent distribution (which is pre-determined). This model has been used to study agents with bounded rationality. (iii) \emph{MWU} workers \cite{littlestone1994weighted} update their strategies according to the celebrated multiplicative weights update algorithm. This model has been used to study adaptive learning agents. 
Our first set of experiments is a continuation to the last subsection. To be more specific, we first focus on the estimation bias of the data requester's cumulative utility $R$. This value is used as the reward in RIL and is calculated from the estimates of the aggregated label accuracy.
This set of experiments aim to investigate whether our RIL module successfully leverages the label accuracy estimates and picks up the right reward signal. As Figure~\ref{figure:rewardError} shows, the estimates only deviate from the real values in a very limited magnitude after a few episodes of learning, regardless of which worker model the experiments run with. The results further demonstrate that our RIL module observe reliable rewards. The next set of experiments is about how quickly RIL learns. As Figure~\ref{figure:learningCurve} shows, under all three worker models, RIL manages to pick up and stick to a promising policy 
in less than 100 episodes. 
This observation also demonstrates the robustness of RIL under different environments.


Our last set of experiments in this subsection aim to evaluate the competitiveness of the policy learned by RIL. In Table~\ref{table:performance}, we use the policy learned after 500 episodes with exploration rate turned off (i.e. $\epsilon =0$) and compare it with three benchmarks constructed by ourselves. To create the first one, Fixed Optimal, we try all 4 possible fixed value for the scaling factor and report the highest cumulative reward realized by either of them.
To create the second one, Heuristic Optimal, we divide the value region of $\tilde{A}^t$ into five regions: $[0,0.6)$, $[0.6,0.7)$, $[0.7,0.8)$, $[0.8,0.9)$ and $[0.9,1.0]$. For each region, we select a fixed value for the scaling factor $a_t$. We traverse all $4^5=1024$ possible combinations to decide the optimal heuristic strategy.
To create the third one, Adaptive Optimal, we change the scaling factor every $4$ steps and report the highest cumulative reward via traversing all $4^7=16384$ possible configurations. This benchmark is infeasible to be reproduced in real-world practice, once the number of steps becomes large. Yet it is very close to the global optimal in the sequential setting.
As Table 1 demonstrates, the two benchmarks plus RIL all achieve a similar performance tested under rational and QR workers.
This is because these two kinds of workers have a fixed pattern in responding to incentives and thus the optimal policy would be a fixed scaling factor throughout the whole episode.
On contrast, MWU workers adaptively learn utility-maximizing strategies gradually, and the learning process is affected by the incentives. Under this worker environment, RIL managers to achieve an average utility score of $15.6$, which is a significant improvement over fixed optimal and heuristic optimal (which achieve $11.7$ and $12.3$ respectively) considering the unrealistic global optimal is only around $18.5$. Up to this point, with three sets of experiments, we demonstrate the competitiveness of RIL and its robustness under different work environments.
Note that, when constructing the benchmarks, we also conduct experiments on DG13, the state-of-the-art peer prediction mechanism for binary labels~\cite{dasgupta2013crowdsourced}, and get the same conclusion.
For example, when DG13 and MWU workers are tested for Fixed Optimal and Heuristic Optimal, the cumulative utilities are $11.537 (.397)$ and $11.908 (0.210)$, respectively, which also shows a large gap with RIL.

\subsection{Empirical Analysis on One Step Payments}
\begin{figure}[t!]
    \centering
    \begin{subfigure}[t]{0.315\textwidth}
        \centering
        \includegraphics[width=\textwidth]{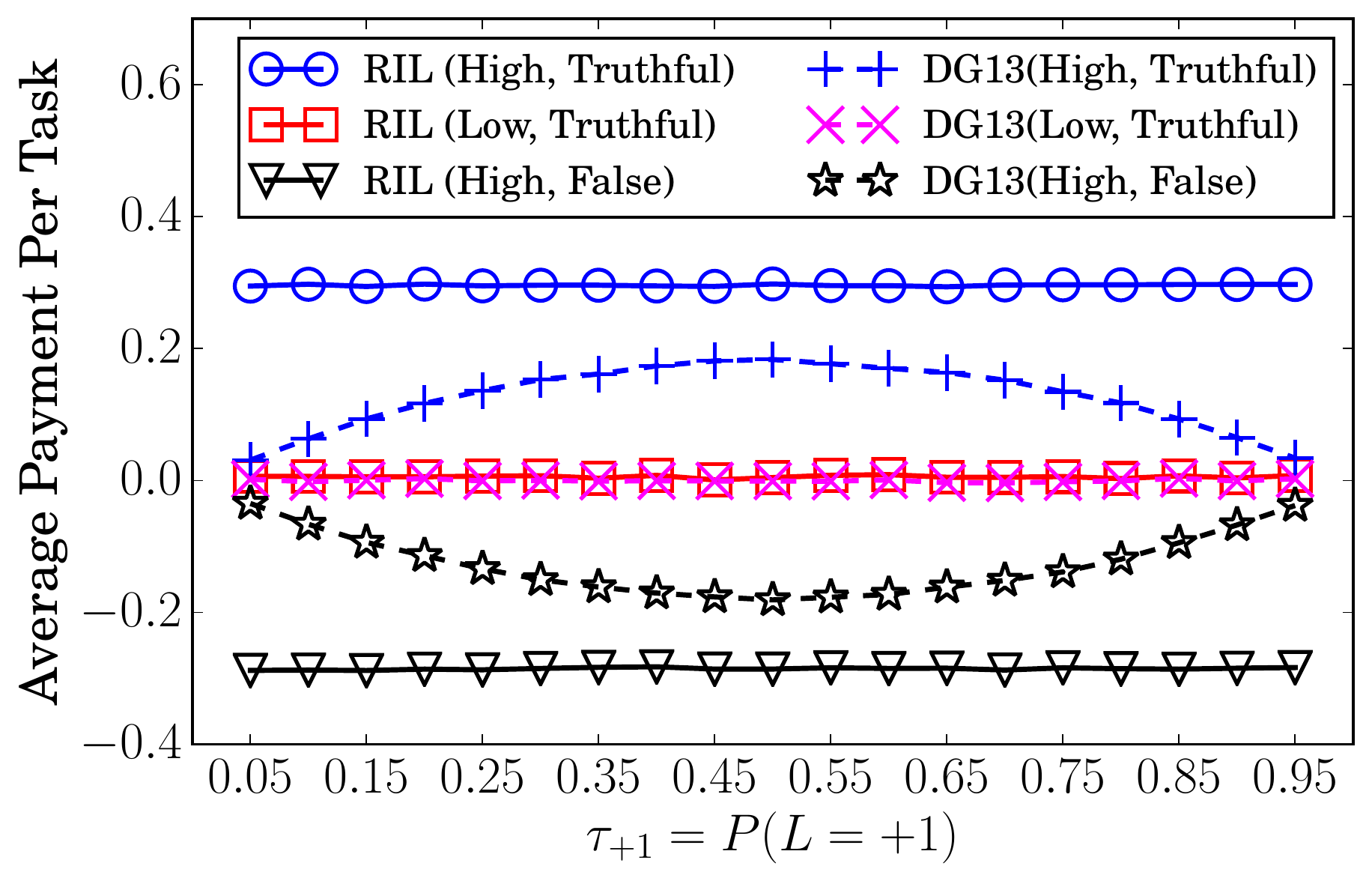}
        \caption{\label{BIM2}}
    \end{subfigure}%
    ~
    \begin{subfigure}[t]{0.325\textwidth}
        \centering
       \includegraphics[width=\textwidth]{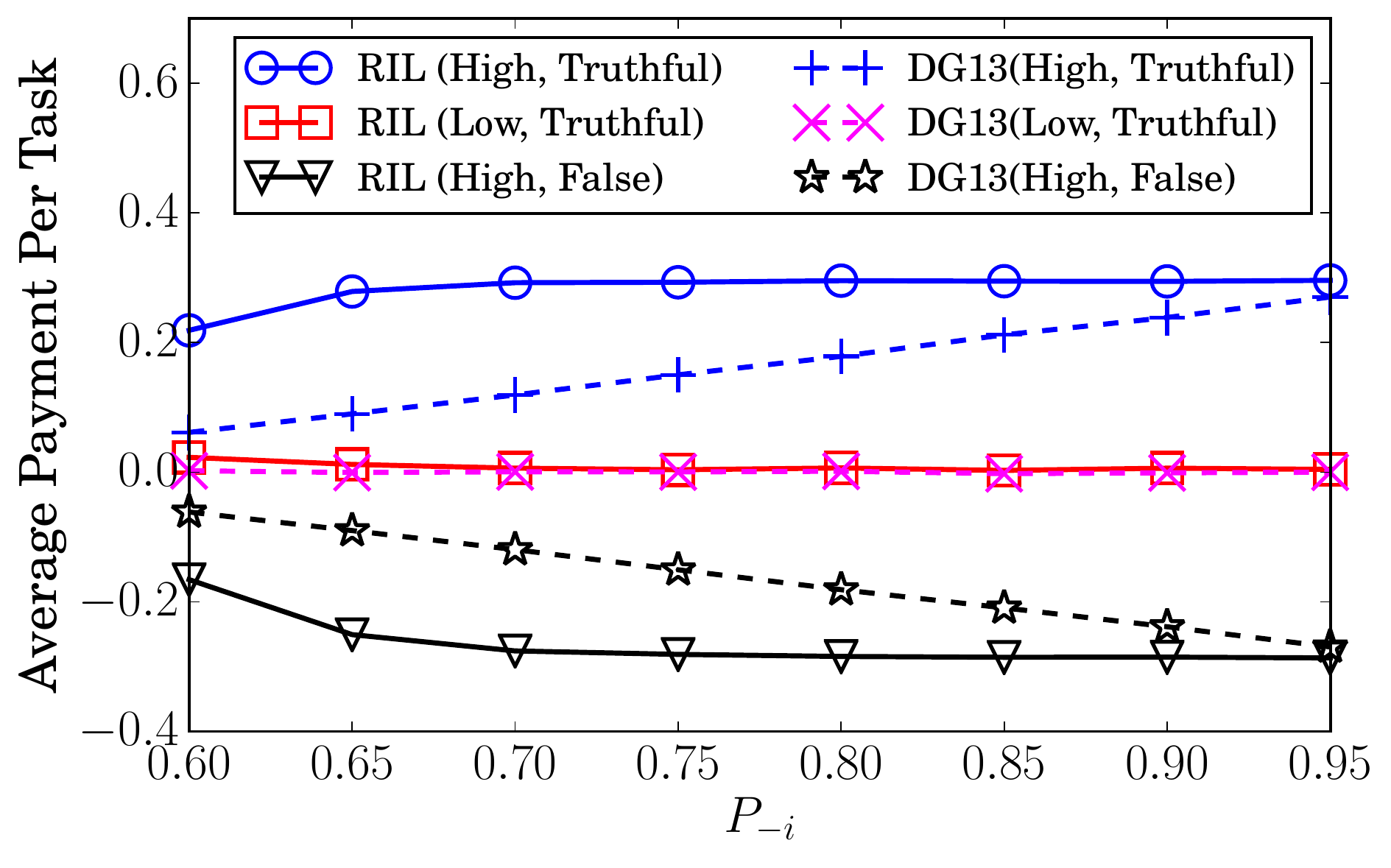}
        \caption{\label{BIM3}}
    \end{subfigure}
        ~
    \begin{subfigure}[t]{0.325\textwidth}
        \centering
        \includegraphics[width=\textwidth]{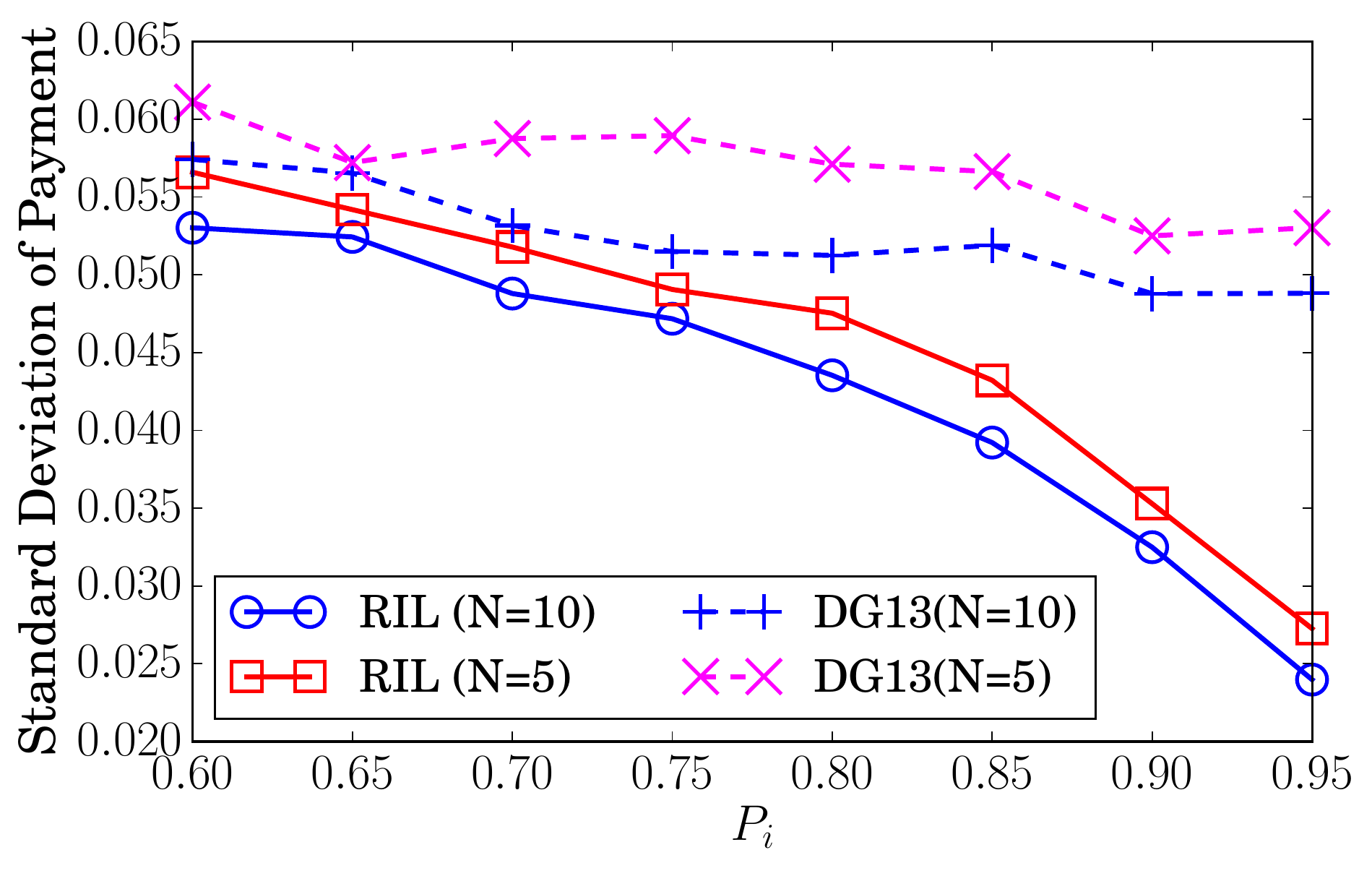}
        \caption{\label{BIM4}}
    \end{subfigure}
    \caption{\label{BIM} Empirical analysis on our Bayesian inference algorithm, averaged over 1000 runs. (a) Average payment per task given true label's distribution. (b) Average payment per task given PoBCs of workers excluding $i$. (c) The standard deviation of the payment given worker $i$'s PoBC.}
    \vspace{-5mm}
\end{figure}

In this subsection, we compare the one step payments provided by our mechanism with the payments calculated by DG13, the state-of-the-art peer prediction mechanism for binary labels~\cite{dasgupta2013crowdsourced}.
We fix the scaling factor $a_t=1$ and set $M=100$, $N=10$, $\mathbb{P}_H=0.8$, $b=0$ and $m_i^t=90$. 
To set up the experiments, we generate task $j$'s true label $\mathcal{L}(j)$ following its distribution $\bm{\tau}$ (to be specified) and worker $i$'s label for task $j$ based on $i$'s PoBC $\mathbb{P}_i$ and $\mathcal{L}(j)$.
In Figure~\ref{BIM2}, we let all workers excluding $i$ report truthfully and exert high efforts (i.e. $\mathbb{P}_{-i} = \mathbb{P}_H$), 
and increase $\tau_{+1}$ from $0.05$ to $0.95$.
In Figure~\ref{BIM3}, we let $\tau_{+1}=0.5$, and increase other workers' PoBCs $\mathbb{P}_{-i}$ from $0.6$ to $0.95$.
As both figures reveal, in our mechanism, the payment for worker $i$ almost only depends on his/her own strategy. On contrast, in DG13, the payments are clearly affected by the distribution of true labels and the strategies of other workers.
In other words, our Bayesian inference is more robust to different environments.
Furthermore, in Figure~\ref{BIM4}, we present the standard deviation of the payment to worker $i$.
We let $\tau_{+1}=0.5$, $\mathbb{P}_{-i}=\mathbb{P}_H$ and increase $\mathbb{P}_i$ from $0.6$ to $0.95$.
As shown in the figure, our method manages to achieve a noticeably smaller standard deviation compared to DG13.
Note that, in Figure~\ref{BIM3}, we implicitly assume that most of workers will at least not adversarially report false labels, which is widely-adopted in previous studies~\cite{liu2012variational}.
For workers' collusion attacks, we also have some defending tricks provided in Appendix F.

\section{Conclusion}
In this paper, we build an inference-aided reinforcement mechanism leveraging Bayesian inference and reinforcement learning techniques to learn the optimal policy to incentivize high-quality labels from crowdsourcing. Our mechanism is proved to be incentive compatible. Empirically, we show that our Bayesian inference algorithm can help improve the robustness and lower the variance of payments, which are favorable properties in practice.
Meanwhile, our reinforcement incentive learning (RIL) algorithm ensures our mechanism to perform consistently well under different worker models.

\section*{Acknowledgments}
This work was conducted within Rolls-Royce@NTU Corporate Lab
with support from the National Research Foundation (NRF) Singapore
under the Corp Lab@University Scheme. Yitao is partially supported by NSF grants \#IIS-1657613, \#IIS-1633857 and DARPA XAI grant \#N66001-17-2-4032. The authors also thank Anxiang Zeng from Alibaba Group for valuable discussions.
\bibliographystyle{plain}
\bibliography{ref}

\renewcommand{\thesection}{\Alph{section}}
\setcounter{section}{0}
\section*{Appendix}
\section{Derivation of Posterior Distribution}
It is not had to figure out the joint distribution of the collected labels $\bm{L}$ and the true labels $\mathcal{L}$ 
\begin{equation}
\label{JointDist}
\begin{split}
    \mathbb{P}(\bm{L}, \mathcal{L}| \bm{\theta}, \bm{\tau})={\prod}_{j=1}^{M}{\prod}_{k\in \{-1,+1\}}\left\{\tau_{k}\prod_{i=1}^{N}\mathbb{P}_i^{\delta_{ijk}}(1-\mathbb{P}_i)^{\delta_{ij(-k)}} \right\}^{\xi_{jk}}
\end{split}
\end{equation}
where $\bm{\theta}=[\mathbb{P}_1,\ldots, \mathbb{P}_N]$ and $\bm{\tau}=[\tau_{-1},\tau_{+1}]$. $\tau_{-1}$ and $\tau_{+1}$ denote the distribution of true label $-1$ and $+1$, respectively.
Besides,  $\delta_{ijk}=\mathbbm{1}(L_i(j)=k)$ and $\xi_{jk}= \mathbbm{1}(\mathcal{L}(j)=k)$.
Then, the joint distribution of $\bm{L}$, $\mathcal{L}$, $\bm{\theta}$ and $\bm{\tau}$ 
\begin{equation}
\label{JointDist2}
\begin{split}
&\mathbb{P}(\mathcal{L},\bm{L},\bm{p}, \bm{\tau})=\mathbb{P}(\mathcal{L},\bm{L}|\bm{p}, \bm{\tau})\cdot \mathbb{P}(\bm{\theta}, \bm{\tau})\\
&=\frac{1}{B(\bm{\beta})}\prod_{k\in \{-1,+1\}}\tau_k^{\hat{\beta}^{*}_k-1}\cdot\prod_{i=1}^{N}\frac{1}{B(\bm{\alpha})}p_i^{\hat{\alpha}^{*}_{i1}-1}(1-p_i)^{\hat{\alpha}^{*}_{i2}-1}
\end{split}
\end{equation}
where $B(x,y)=(x-1)!(y-1)!/(x+y-1)!$ denotes the beta function, and
\begin{equation*}
\begin{split}
&\hat{\alpha}^{*}_{i1}={\sum}_{j=1}^{M}{\sum}_{k=1}^{K}\delta_{ijk}\xi_{jk}+\alpha_{1}\\
&\hat{\alpha}^{*}_{i2}={\sum}_{j=1}^{M}{\sum}_{k=1}^{K}\delta_{ij(3-k)}\xi_{jk}+\alpha_{2}\\
&\hat{\beta}^{*}_k={\sum}_{j=1}^{M}\xi_{jk}+\beta_{k}.
\end{split}
\end{equation*}
In this case, we can conduct marginalization via integrating the joint distribution $\mathbb{P}(\mathcal{L},\bm{L},\bm{p}, \bm{\tau})$ over $\bm{\theta}$ and $\bm{\tau}$ as
\begin{equation}
\label{marginalization}
\begin{split}
P(\mathcal{L},\bm{L}|\bm{\alpha}, \bm{\beta})=\frac{B(\hat{\bm{\beta}})}{B(\bm{\beta})}\cdot {\prod}_{i=1}^{N}\frac{B(\hat{\bm{\alpha}}_{i})}{[B(\bm{\alpha})]^2}
\end{split}
\end{equation}
where $\hat{\bm{\alpha}}_i=[\hat{\alpha}^{*}_{i1}+\alpha_1-1,\hat{\alpha}^{*}_{i2}+\alpha_2-1]$ and $\hat{\bm{\beta}}=[\hat{\beta}^{*}_{-1}+\beta_{-1}-1,\hat{\beta}^{*}_{+1}+\beta_{+1}-1]$. Following Bayes' theorem, we can know that
\begin{equation}
\label{PostDist}
P(\bm{L}|\mathcal{L})=\frac{P(\mathcal{L},\bm{L}|\bm{\alpha}, \bm{\beta})}{P(\mathcal{L}|\bm{\alpha}, \bm{\beta})}\propto B(\hat{\bm{\beta}})\prod_{i=1}^{N}B(\hat{\bm{\alpha}}_{i}). 
\end{equation}
\section{Proof for Lemma 1}
\subsection{Basic Lemmas}
We firstly present some lemmas for our proof later.
\begin{lemma}
\label{MoGene}
If $x\sim \mathrm{Bin}(n,p)$, $\mathbb{E}t^x= \left(1-p+tp\right)^{n}$ holds for any $t>0$, where $\mathrm{Bin}(\cdot)$ is the binomial distribution.
\begin{proof}
\begin{equation}
t^x = e^{x\log t}=m_x(\log t)= \left(1-p+pe^{\log t}\right)^{n}
\end{equation}
where $m_x(\cdot)$ denotes the moment generating function.
\end{proof}
\end{lemma}

\begin{lemma}
\label{SolveF}
For given $n,m\geq 0$, if $0\leq p\leq 1$, we can have
\begin{equation*}
\begin{split}
&{\sum}_{x=0}^{n}{\sum}_{w=0}^{m} C_{n}^{x}C_{m}^{w}p^{x+w}(1-p)^{y+z}\times\\
&\qquad\qquad\qquad B(x+z+1+t,y+w+1)=\\
&\int_{0}^{1}[(2p-1)x+1-p]^{n}[(1-2p)x+p]^{m}x^{t}\mathrm{d}x
\end{split}
\end{equation*}
\begin{proof}
By the definition of the beta function~\cite{olver2010nist},
\begin{equation}
B(x, y) = \int_{0}^{+\infty} u^{x-1}(1+u)^{-(x+y)}\mathrm{d}u
\end{equation}
we can have
\begin{align}
&\sum_{x,w} C_{n}^{x}C_{m}^{w}p^{x+w}(1-p)^{y+z}B(x+z+1+t,y+w+1)\nonumber\\
&= \int_{0}^{+\infty} \mathbb{E}u^{x}\cdot\mathbb{E}u^z \cdot u^t\cdot (1+u)^{-(n+m+2+t)}\mathrm{d}u
\end{align}
where we regard $x\sim \mathrm{Bin}(n,p)$ and $z\sim \mathrm{Bin}(m,1-p)$.
Thus, according to Lemma~\ref{MoGene}, we can obtain
\begin{equation}
\begin{split}
&\int_{0}^{+\infty} \mathbb{E}u^{x}\cdot\mathbb{E}u^z \cdot u^t\cdot (1+u)^{-(n+m+3)}\mathrm{d}u\\
&=\int_{0}^{+\infty} \frac{[1-p+up]^n\cdot [p+(1-p)u]^m\cdot u^t}{(1+u)^{n+m+2+t}}\mathrm{d}u.
\end{split}
\end{equation}
For the integral operation, substituting $u$ with $v-1$ at first and then $v$ with $(1-x)^{-1}$, we can conclude Lemma~\ref{SolveF}.
\end{proof}
\end{lemma}

\begin{lemma}
\label{Sum1}
$\sum_{n=0}^{N}C_N^{n}\cdot x^{n}=(1+x)^N$.
\end{lemma}
\begin{lemma}
\label{Sum2}
$\sum_{n=0}^{N} C_N^{n}\cdot n\cdot x^{n}=N\cdot x\cdot (1+x)^{N-1}$.
\end{lemma}
\begin{lemma}
\label{Sum3}
$\sum_{n=0}^{N} C_N^{n}\cdot n\cdot x^{N-n}=N\cdot (1+x)^{N-1}$.
\end{lemma}
\begin{lemma}
\label{Sum5}
$\sum_{n=0}^{N} C_N^{n}\cdot n^2\cdot x^{n}=Nx(1+Nx)(1+x)^{N-2}$.
\end{lemma}
\begin{lemma}
\label{Sum6}
$\sum_{n=0}^{N} C_N^{n}\cdot n^2\cdot x^{N-n}=N(x+N)(1+x)^{N-2}$.
\end{lemma}
\begin{lemma}
\label{Sum4}
If $0<x<1$, we can have
\begin{equation*}
\begin{split}
\sum_{n=0}^{\lfloor N/2\rfloor} C_N^{n}\cdot x^{n} &\geq \left(1-e^{-cN}\right) \cdot (1+x)^{N}\\
\sum_{n=\lfloor N/2 \rfloor +1}^{N} C_N^{n}\cdot x^{N-n}&\geq \left(1-e^{-cN}\right) \cdot (1+x)^{N}.
\end{split}
\end{equation*}
where $c=0.5(1-x)^2(1+x)^{-2}$.
\begin{proof}
To prove the lemmas above, we firstly define
\begin{equation}
F_t(x)=\sum_{n=0}^{N} C_N^{n} n^t x^{n}
\end{equation}
Then, Lemma~\ref{Sum1} can be obtained by expanding $(1+x)^N$.
Lemma~\ref{Sum2} can be proved as follows
\begin{equation}
\begin{split}
F_1(x)&=\sum_{n=0}^{N} C_N^{n} (n+1) x^{n}-(1+x)^N\\
\sum_{n=0}^{N} C_N^{n} (n+1) x^{n}&=\frac{\mathrm{d}}{\mathrm{d}x}\left[xF_0(x)\right]\\
&=Nx(1+x)^{N-1}+(1+x)^N.
\end{split}
\end{equation}
Lemma~\ref{Sum3} can be obtained as follows
\begin{equation}
\begin{split}
\sum_{n=0}^{N} C_N^{n} n x^{N-n}&=x^N\sum_{n=0}^{N} C_N^{n} n \left(\frac{1}{x}\right)^n\\
&=x^N\cdot N\cdot \frac{1}{x}\cdot \left(1+\frac{1}{x}\right)^{N-1}.
\end{split}
\end{equation}
For Lemma~\ref{Sum5}, we can have
\begin{align}
F_2(x)&=\sum_{n=0}^{N} C_N^{n} (n+2)(n+1) x^{n}-3F_{1}(x)-2F_{0}(x)\nonumber \\
&=\left[x^2F_0(x)\right]' -3F_{1}(x)-2F_{0}(x)
\end{align}
Thus, we can have
\begin{equation}
F_2(x)=Nx(1+Nx)(1+x)^{N-2}
\end{equation}
which concludes Lemma~\ref{Sum5}.
Then, Lemma~\ref{Sum6} can be obtained by considering Eqn. (\ref{Eqv}).
\begin{equation}
\label{Eqv}
\begin{split}
\sum_{n=0}^{N} C_N^{n} n^2 x^{N-n}=x^N\sum_{n=0}^{N} C_N^{n} n^2 \left(\frac{1}{x}\right)^n.
\end{split}
\end{equation}
For Lemma~\ref{Sum4}, we can have
\begin{equation}
\sum_{n=0}^{\lfloor N/2\rfloor} C_N^{n} x^{n}=(1+x)^{N}\sum_{n=0}^{\lfloor N/2\rfloor} C_N^{n} p^n (1-p)^{N-n}
\end{equation}
where $p=x(1+x)^{-1}$. Let $X\sim \mathrm{Bin}(N, p)$, we can have
\begin{equation}
\sum_{n=0}^{\lfloor N/2\rfloor} C_N^{n} p^n (1-p)^{N-n}\geq 1-P\left(X\geq N/2\right).
\end{equation}
Since $x<1$, $p<0.5$ and $Np<N/2$. Considering Hoeffding's inequality, we can get
\begin{equation}
P\left(X\geq N/2\right)\leq \exp \left[-\frac{N(1-x)^2}{2(1+x)^2}\right]
\end{equation}
which concludes the first inequality in Lemma~\ref{Sum4}. Similarly, for the second inequality, we can have
\begin{equation}
\sum_{n=K}^{N} C_N^{n}x^{N-n}=(1+x)^{N}\sum_{n=K}^{N} C_N^{n} (1-p)^n p^{N-n}
\end{equation}
where $K=\lfloor N/2 \rfloor +1$. Suppose $Y\sim \mathrm{Bin}(N, 1-p)$, we can have
\begin{equation}
\sum_{n=K}^{N} C_N^{n} (1-p)^n p^{N-n}\geq 1-P\left(Y\leq N/2\right).
\end{equation}
Considering Hoeffding's inequality, we can also get
\begin{equation}
P\left(Y\leq N/2\right)\leq \exp \left[-\frac{N(1-x)^2}{2(1+x)^2}\right]
\end{equation}
which concludes the second inequality in Lemma~\ref{Sum4}.
\end{proof}
\end{lemma}

\begin{lemma}
\label{Inequality1}
For any $x,y\geq 0$, we can have
$$(1+x)^{y}\leq e^{xy}.$$
\begin{proof}
Firstly, we can know $(1+x)^{y}=e^{y\log(1+x)}$. Let $f(x)=x-\log(x)$. Then, we can have $f(0)=0$ and $f'(x)\geq 0$. Thus, $x\geq\log (1+x)$ and we can conclude Lemma~\ref{Inequality1} by taking this inequality into the equality.
\end{proof}
\end{lemma}

\begin{lemma}
\label{Concave1}
$$g(x)=\frac{e^x}{e^x+1}$$
is a concave function when $x\in [0,+\infty)$.
\begin{proof}
$g'(x)= (2+t(x))^{-1}$, where $t(x)=e^x+e^{-x}$. $t'(x)=e^x-e^{-x}\geq 0$ when $x\in [0,+\infty)$. Thus, $g'(x)$ is monotonically decreasing when $x\in [0,+\infty)$, which concludes Lemma~\ref{Concave1}.
\end{proof}
\end{lemma}

\begin{lemma}
\label{Concave2}
For $x\in(-\infty, +\infty)$, 
$$h(x)=\frac{1}{e^{|x|}+1}$$
satisfies
$$h(x)< e^x \;\;\mathrm{and}\;\; h(x)< e^{-x}.$$
\begin{proof}
When $x\geq 0$, we can have
\begin{equation}
h(x)<\frac{1}{e^x}=e^{-x}\leq e^{x}.
\end{equation}
When $x\leq 0$, we can have
\begin{equation}
h(x)=\frac{e^x}{e^x+1}<e^{x}\leq e^{-x}.
\end{equation}
\end{proof}
\end{lemma}

\begin{lemma}
\label{Concave3}
If $\lambda=p/(1-p)$ and $0.5<p<1$, then
$${\sum}_{n=\lfloor N/2 \rfloor}^{N}C_N^n\lambda^{m-n}p^n(1-p)^m\leq [4p(1-p)]^{N/2}$$
$${\sum}_{n=0}^{\lfloor N/2 \rfloor}C_N^n\lambda^{n-m}p^n(1-p)^m\leq [4p(1-p)]^{N/2}$$
where $m=N-n$.
\begin{proof}
For the first inequality, we can have
\begin{align}
&\sum_{n=\lfloor N/2 \rfloor}^{N}C_N^n\lambda^{m-n}p^n(1-p)^m \\&=\sum_{n=\lfloor N/2 \rfloor}^{N}C_N^np^m(1-p)^n\leq \sum_{m=0}^{\lfloor N/2 \rfloor}C_N^m p^m(1-p)^n\nonumber
\end{align}
According to the inequality in \cite{Arratia1989}, we can have
\begin{equation}
\label{xxx}
 \sum_{m=0}^{\lfloor N/2 \rfloor}C_N^m p^m(1-p)^n \leq \exp (-ND)
\end{equation}
where $D=-0.5\log(2p)-0.5\log(2(p-1))$, which concludes the first inequality in Lemma~\ref{Concave3}.

For the second inequality, we can have
\begin{equation}
\begin{split}
&\sum_{n=0}^{\lfloor N/2 \rfloor}C_N^n\lambda^{n-m}p^n(1-p)^m \\
&= \frac{1}{[p(1-p)]^N}\sum_{n=0}^{\lfloor N/2 \rfloor}C_N^n[p^3]^n[(1-p)^3]^m
 \\&=\frac{[p^3+(1-p)^3]^N}{[p(1-p)]^N}\sum_{n=0}^{\lfloor N/2 \rfloor}C_N^n x^n(1-x)^m
\end{split}
\end{equation}
where $x=p^3/[p^3+(1-p)^3]$. By using Eqn. (\ref{xxx}), we can have
\begin{equation}
\begin{split}
&\sum_{n=0}^{\lfloor N/2 \rfloor}C_N^n\lambda^{n-m}p^n(1-p)^m\\
&\leq \frac{[p^3+(1-p)^3]^N}{[p(1-p)]^N} [x(1-x)]^{N/2} \\
&=[4p(1-p)]^{N/2}
\end{split}
\end{equation}
which concludes the second inequality of Lemma~\ref{Concave3}.
\end{proof}
\end{lemma}

\subsection{Main Proof}
To prove Lemma 1, we need to analyze the posterior distribution of $\mathcal{L}$ which satisfies
\begin{equation}
\label{postdist2}
\mathbb P(\mathcal{L}|\bm{L})=B(\hat{\bm{\beta}}){\prod}_{i=1}^{N}B(\hat{\bm{\alpha}}_{i})/[C_p\cdot \mathbb P(\bm{L})]
\end{equation}
where $C_p$ is the nomalization constant.
This is because the samples are generated based on this distribution.
However, both the numerator and denominator in Eqn.~(\ref{postdist2}) are changing with $\bm{L}$, making the distribution difficult to analyze.
Thus, we derive a proper approximation for the denominator of Eqn. (\ref{postdist2}) at first.
Denote the labels generated by $N$ workers for task $j$ as vector $\bm{L}(j)$.
The distribution of $\bm{L}(j)$ satisfies
\begin{equation}
\mathbb{P}_{\hat{\bm{\theta}}}[\bm{L}(j)] = {\sum}_{k\in\{-1,+1\}}\tau_k{\prod}_{i=1}^{N}\mathbb{P}_i^{\delta_{ijk}}(1-\mathbb{P}_i)^{\delta_{ij(-k)}}
\end{equation}
where $\hat{\bm{\theta}}=[\tau_{-1}, \mathbb{P}_1,\ldots,\mathbb{P}_N]$ denotes all the parameters and $\delta_{ijk}=\mathbbm{1}(L_i(j)=k)$.
Then, we can have
\begin{lemma}
\label{Denominator}
When $M\rightarrow \infty$, 
\begin{equation*}
\mathbb{P}(\bm{L})\rightarrow C_{L}(M) \cdot {\prod}_{\bm{L}(j)} \left\{\mathbb{P}_{\hat{\bm{\theta}}}[\bm{L}(j)]\right\}^{M\cdot \mathbb{P}_{\hat{\bm{\theta}}}[\bm{L}(j)]}
\end{equation*}
where $C_{L}(M)$ denotes a constant that depends on $M$.
\begin{proof}
Denote the prior distribution of $\bm{\theta}$ by $\pi$. Then,
\begin{align}
&P(\mathcal{L}|\bm{\alpha}, \bm{\beta})= {\prod}_{j=1}^{M}P_{\bm{\theta}}(\bm{x}_j) \int e^{[-M\cdot d_{KL}]} \mathrm{d}\pi(\hat{\bm{\theta}})\\
&d_{KL}=\frac{1}{M}\sum_{j=1}^{M}\log \frac{P_{\bm{\theta}}(\bm{x}_j)}{P_{\bm{\hat{\theta}}}(\bm{x}_j)}\rightarrow \mathrm{KL}[P_{\bm{\theta}}(\bm{x}),P_{\bm{\hat{\theta}}}(\bm{x})]
\end{align}
where $\bm{x}_j$ denotes the labels generated for task $j$. The KL divergence $\mathrm{KL}[\cdot, \cdot]$, which denotes the expectation of the log-ratio between two probability distributions, is a constant for the given $\bm{\theta}$ and $\hat{\bm{\theta}}$.
Thus, $\int e^{[-M\cdot d_{KL}]} \mathrm{d}\pi(\hat{\bm{\theta}})=C_{L}(M)$.
In addition, when $M\rightarrow \infty$, we can also have $\sum 1(\bm{x}_j=\bm{x})\rightarrow M \cdot P_{\bm{\theta}}(\bm{x})$, which concludes Lemma~\ref{Denominator}.
\end{proof}
\end{lemma}

Then, we move our focus back to the samples.
To quantify the effects of the collected labels, we introduce a set of variables to describe the real true labels and the collected labels.
Among the $n$ tasks of which the posterior true label is correct,
\begin{itemize}[noitemsep,topsep=0pt]
\item $x_0$ and $y_0$ denote the number of tasks of which the real true label is $-1$ and $+1$, respectively.
\item $x_i$ and $y_i$ denote the number of tasks of which worker $i$'s label is correct and wrong, respectively.
\end{itemize}
Also, among the remaining $m=M-n$ tasks, 
\begin{itemize}[noitemsep,topsep=0pt]
\item $w_0$ and $z_0$ denote the number of tasks of which the real true label is $-1$ and $+1$, respectively.
\item $w_i$ and $z_i$ denote the number of tasks of which worker $i$'s label is correct and wrong, respectively.
\end{itemize}
Thus, we can have $x_i+y_i=n$ and $w_i+z_i=m$. Besides, we use $\xi_i$ to denote the combination $(x_i,y_i,w_i, z_i)$.

To compute the expectation of $m/M$, we need to analyze the probability distribution of $m$. According to Eqn.~(\ref{PostDist}), we can know that $\mathbb{P}(m)$ satisfies
\begin{equation}
\mathbb{P}(m) \approx \frac{C_{M}^{m}}{Z} \sum_{\xi_0,\ldots, \xi_N}\prod_{i=0}^{N}\mathbb{P}(\xi_i|m) B(\hat{\bm{\beta}})\prod_{i=1}^{N}B(\hat{\bm{\alpha}}_{i})
\end{equation}
where $Z=C_pC_L{\prod}_{\bm{x}} [P_{\bm{\theta}}(\bm{x})]^{M\cdot P_{\bm{\theta}}(\bm{x})}$ is independent of $\xi_i$ and $m$.
Meanwhile, $\hat{\beta}_{-1}=x_0+z_0+1$, $\hat{\beta}_{+1}=y_0+w_0+1$, $\hat{\alpha}_{i1}=x_i+z_i+2$ and $\hat{\alpha}_{i2}=x_i+z_i+1$.
When the $m$ tasks of which the posterior true label is wrong are given, we can know that $x_i\sim \mathrm{Bin}(n, \mathbb{P}_i)$ and $w_i\sim \mathrm{Bin}(m, \mathbb{P}_i)$, where $\mathrm{Bin}(\cdot)$ denotes the binomial distribution.
In addition, $x_i$ and $y_i$ are independent of $w_i$, $z_i$ and $\xi_{k\neq i}$.
Also, $w_i$ and $z_i$ are independent of $x_i$ and $y_i$ and $\xi_{k\neq i}$.
Thus, we can further obtain $\mathbb{P}(m)\approx \hat{Z}^{-1}\cdot C_{M}^{m}Y(m)$, where
\begin{equation}
\label{PDist}
\begin{split}
&Y(m) =e^{\log H(m,\mathbb{P}_0;M,0)+\sum_{i=1}^{N}\log H(m,\mathbb{P}_i;M,1)}\\
&H(m,p;M,t)={\sum}_{x=0}^{n}{\sum}_{w=0}^{m} 2^{M+1}C_{n}^{x}C_{m}^{w}\times\\
&\quad p^{x+w}(1-p)^{y+z}B(x+z+1+t,y+w+1)
\end{split}
\end{equation}
and $\hat{Z}=2^{-(N+1)(M+1)}Z$.
Considering $\sum_{m=1}^{M} \mathbb{P}(m)=1$, we can know that $\hat{Z}\approx{\sum}_{m=1}^{M}C_{M}^{m}Y(m)$. Note that, we use $\mathbb{P}_0$ to denote the probability of true label $1$, namely $\tau_1$.

\textbf{The biggest challenge of our proof} exists in analyzing function $H(m,p;M,t)$ which we put in the next subsection (Section C.3).
Here, we directly use the obtained lower and upper bounds  depicted in Lemmas~\ref{LowBound1} and \ref{UpBound1} and can have
\begin{equation}
\left\{
\begin{array}{lc}
e^{C-{K}_l m}\lesssim Y(m) \lesssim e^{C-K_u m} & 2m\leq M\\
e^{C+\delta-{K}_l  n}\lesssim Y(m) \lesssim e^{C+\delta-K_u n} & 2m>M
\end{array}
\right.
\end{equation}
where $C=H(0,\mathbb{P}_0;M,0)+\sum_{i=1}^{N}H(0,\mathbb{P}_i;M,1)$ and
\begin{equation*}
\begin{split}
&K_l = {\sum}_{i=0}^{N}\log \hat{\lambda}_{i}\;,\; K_u =  2{\sum}_{i=0}^{N}\log \left(2\hat{\mathbb{P}}_i\right)\\
&\delta = \Delta\cdot \log(M)+{\sum}_{i=1}^{N}(-1)^{1(\mathbb{P}_i>0.5)}\phi(\hat{\mathbb{P}}_i)\\
&\hat{\lambda}_i=\max\left\{\frac{\mathbb{P}_i}{\bar{\mathbb{P}}_i+\frac{1}{M}},\frac{\bar{\mathbb{P}}_i}{\mathbb{P}_i+\frac{1}{M}}\right\}
\;,\;\phi (p) =\log\frac{2\mathbb{P}-1}{\mathbb{P}}\\
&\Delta={\sum}_{i=1}^N[1(\mathbb{P}_i<0.5)-1(\mathbb{P}_i>0.5)].
\end{split}
\end{equation*}
Here, $\bar{\mathbb{P}}=1-\mathbb{P}$, $\hat{\mathbb{P}}=\max \{\mathbb{P}, \bar{\mathbb{P}}\}$ and $\mathbb{P}_0=\tau_{-1}$. Besides, we set a convention that $\phi(p)=0$ when $p=0.5$. Thereby, the expectations of $m$ and $m^2$ satisfy
\begin{align}
\mathbb{E}[m] \lesssim \frac{\sum_{m=0}^{M}me^{-K_u m}+\sum_{m=0}^{M}me^{\delta-K_u n}}{\sum_{m=0}^{k}e^{-K_l m}+\sum_{m=k+1}^{M}e^{\delta-K_l n}}
\label{ExM1}\\
\mathbb{E}[m^2] \lesssim \frac{\sum_{m=0}^{M}m^2e^{-K_u m}+\sum_{m=0}^{M}m^2e^{\delta-K_u n}}{\sum_{m=0}^{k}e^{-K_l m}+\sum_{m=k+1}^{M}e^{\delta-K_l n}} \label{ExM2}
\end{align}
where $k=\lfloor M/2 \rfloor$.
By using Lemmas~\ref{Sum2}, \ref{Sum3}, \ref{Sum5} and \ref{Sum6}, we can know the upper bounds of the numerator in Eqn.~(\ref{ExM1}) and (\ref{ExM2}) are $M(\varepsilon+e^{\delta})(1+\varepsilon)^{M-1}$ and $[M^2\varepsilon^2+M\varepsilon+e^{\delta}(M^2+M\varepsilon)](1+\varepsilon)^{M-2}$, respectively, where $\varepsilon=e^{-K_u}$. On the other hand, by using Lemma~\ref{Sum4}, we can obtain the lower bound of the denominator as $(1+e^{\delta})[1-e^{-c(\omega)M}](1+\omega)^{M}$, where $\omega=e^{-K_l}$ and $c(\omega)=0.5(1-\omega)^2(1+\omega)^{-2}$.
Considering $M\gg 1$, we can make the approximation that $e^{-c(\omega)M}\approx 0$ and $(1+e^\delta)\varepsilon/M\approx 0$. Besides, $(1+\omega)^{M}\geq 1$ holds because $\omega\geq 0$. In this case, Lemma~1 can be concluded by combining the upper bound of the numerator and the lower bound of the denominator.

\subsection{H function analysis}
Here, we present our analysis on the $H$ function defined in Eqn. (\ref{PDist}).
Firstly, we can have:
\begin{lemma}
\label{MidPoint}
$H(m,0.5;M,t) = 2(t+1)^{-1}$.
\end{lemma}
\begin{lemma}
\label{Symmetry}
$H(m,p;M,t) = H(n, \bar{p};M,t)$.
\end{lemma}
\begin{lemma}
\label{LogConvexity}
As a function of $m$, $H(m,p;M,t)$ is logarithmically convex.
\begin{proof}
Lemma~\ref{MidPoint} can be proved by integrating $2 x^t$ on $[0,1]$.
Lemma~\ref{Symmetry} can be proved by showing that $H(n, \bar{p};M,t)$ has the same expression as $H(m,p;M,t)$.
Thus, in the following proof, we focus on Lemma~\ref{LogConvexity}. Fixing $p$, $M$ and $t$, we denote $\log (H)$ by $f(m)$. Then, we compute the first-order derivative as
\begin{equation}
H(m)f'(m)=2^{M+1}\int_{0}^{1}\lambda u^{n}(1-u)^{m}x^t\mathrm{d}x
\end{equation}
where $u= (2p-1)x+1-p$ and $\lambda=\log(1-u)-\log(u)$. Furthermore, we can solve the second-order derivative as
\begin{equation}
\begin{split}
&2^{-2(M+1)}H^2(m)f''(m)=\\
&\int_{0}^{1}g^2(x)\mathrm{d}x\int_{0}^{1}h^2(x)\mathrm{d}x-\left(\int_{0}^{1}g(x)h(x)\mathrm{d}x\right)^2
\end{split}
\end{equation}
where the functions $g,h:(0,1)\rightarrow  \mathbb{R}$ are defined by
\begin{equation}
g=\lambda\sqrt{u^{n}(1-u)^{m}}\;, \; h = \sqrt{u^{n}(1-u)^{m}}.
\end{equation}
By the Cauchy-Schwarz inequality,
\begin{equation}
\int_{0}^{1}g^2(x)\mathrm{d}x\int_{0}^{1}h^2(x)\mathrm{d}x\geq \left(\int_{0}^{1}g(x)h(x)\mathrm{d}x\right)^2
\end{equation}
we can know that $f''(m)\geq 0$ always holds, which concludes that $f$ is convex and $H$ is logarithmically convex.
\end{proof}
\end{lemma}
Then, for the case that $t=1$ and $M\gg 1$, we can further derive the following three lemmas for $H(m,p;M,1)$:
\begin{lemma}
\label{Ratio1}
The ratio between two ends satisfies
$$\log\frac{H(0,p;M,1)}{H(M,p;M,1)} \approx \left\{
    \begin{array}{cl}
    \log(M)+\epsilon(p) & p>0.5\\
    0 & p=0.5\\
    -\log(M)-\epsilon(\bar{p}) & p<0.5
    \end{array}\right.$$
\end{lemma}
where $\epsilon(p)=\log(2p-1)-\log(p)$ and $\epsilon(p)=0$ if $p=0.5$.
\begin{lemma}
\label{LowBound1}
The lower bound can be calculated as
\begin{equation*}
\log H(m,p)\gtrsim \left\{
    \begin{array}{cl}
    H(0,p)- k_l \cdot m& 2m\leq M\\
    H(M,p)- k_l \cdot n& 2m>M
    \end{array}\right.
\end{equation*}
where $k_l=\log\left(\max\left\{p/(\bar{p}+M^{-1}),\bar{p}/(p+M^{-1})\right\}\right)$.
\end{lemma}
\begin{lemma}
\label{UpBound1}
The upper bound can be calculated as
\begin{equation*}
\log H(m,p)\lesssim \left\{
    \begin{array}{cl}
    H(0,p)- k_u\cdot m& 2m\leq M\\
    H(M,p)- k_u\cdot n& 2m>M
    \end{array}\right.
\end{equation*}
where $n=M-m$ and $k_u=2\log\left(2\cdot \max\left\{p,\bar{p}\right\}\right)$.
\begin{proof}
By Lemma~\ref{MidPoint}, $\log H(m, 0.5;M,1)\equiv 0$, which proves the above three lemmas for the case that $p=0.5$. Considering the symmetry ensured by Lemma~\ref{Symmetry}, we thus focus on the case that $p>0.5$ in the following proof and transform $H(m,p)$ into the following formulation
\begin{equation}
H(m,p)=\omega(p)\cdot \int_{\bar{p}}^{p}x^n(1-x)^m(x-1+p)\mathrm{d}x
\end{equation}
where $\omega(p)=2^{M+1}/(2p-1)^2$. Then, we can solve $H(0,p)$ and $H(M,p)$ as
\begin{equation}
\begin{split}
H(0,p)&=\omega(p)\int_{\bar{p}}^{p}x^M(x-\bar{p})\mathrm{d}x\\
&=\frac{(2p)^{M+1}}{(2p-1)(M+1)} - O\left(\frac{(2p)^{M+1}}{M^2}\right)
\end{split}
\end{equation}
\begin{equation}
\begin{split}
&H(M,p)=\omega(p)\int_{\bar{p}}^{p}(1-x)^M(x-\bar{p})\mathrm{d}x\\
&=\frac{p(2p)^{M+1}}{(2p-1)^2(M+1)(M+2)} - O\left(\frac{(2\bar{p})^{M+1}}{M+2}\right).
\end{split}
\end{equation}
Using the Taylor expansion of function $\log(x)$, we can calculate the ratio in Lemma~\ref{Ratio1} as
\begin{equation}
\log\frac{H(0,p)}{H(M,p)}=\log(M)+\log\frac{2p-1}{p}+O\left(\frac{1}{M}\right)
\end{equation}
which concludes Lemma~\ref{Ratio1} when $M\gg 1$.

Furthermore, we can solve $H(1,p)$ as
\begin{equation}
\begin{split}
H(1,p)&= \omega(p)\int_{\bar{p}}^{p}x^{M-1}(x-\bar{p})\mathrm{d}x-H(0,p)\\
&=\frac{(2\bar{p}+M^{-1})(2p)^{M}}{(2p-1)(M+1)} - O\left(\frac{(2p)^{M+1}}{M^2}\right)
\end{split}
\end{equation}
The value ratio between $m=0$ and $m=1$ then satisfies
\begin{equation}
\log \frac{H(1,p)}{H(0,p)}  = \log\frac{p}{\bar{p}+M^{-1}}+O\left(\frac{1}{M}\right).
\end{equation}
By Rolle's theorem, there exists a $c\in [m, m+1]$ satisfying
\begin{equation}
\log H(1,p) - \log H(0,p) = f'(c)
\end{equation}
where $f(m)=\log H(m,p)$. Meanwhile, Lemma~\ref{LogConvexity} ensures that $f''(m)\geq 0$ always holds. Thus, we can have
\begin{equation}
\log H(m+1,p) - \log H(m,p)\geq \log \frac{H(1,0)}{H(0,p)}
\end{equation}
which concludes the first case of Lemma~\ref{LowBound1}. Similarly, we compute the ratio between $m=M-1$ and $M$ as
\begin{equation}
\log \frac{H(M,p)}{H(M-1,p)}  = \log\frac{p}{\bar{p}+M^{-1}}+O\left(\frac{1}{M}\right).
\end{equation}
Meanwhile, Rolle's theorem and Lemma~\ref{LogConvexity} ensure that
\begin{equation}
\log H(m,p) - \log H(m-1,p)\leq \log \frac{H(M,0)}{H(M-1,p)}
\end{equation}
which concludes the second case of Lemma~\ref{LowBound1}.

Lastly, we focus on the upper bound described by Lemma~\ref{UpBound1}. According to the inequality of arithmetic and geometric means, $x(1-x)\leq 2^{-2}$ holds for any $x\in [0,1]$. Thus, when $2m\leq M$ (i.e. $n\geq m$), we can have
\begin{equation}
H(m,p)\leq 2^{-2m}\omega(p)\cdot \int_{\bar{p}}^{p}x^{n-m}(x-1+p)\mathrm{d}x
\end{equation}
where the equality only holds when $m=0$.
\begin{equation}
 \int_{\bar{p}}^{p}x^{n-m}(x-1+p)\mathrm{d}x=\frac{(2p-1)p^{\delta}}{\delta}+\frac{\Delta}{\delta(\delta+1)}
\end{equation}
where $\delta=n-m+1$ and $\Delta=\bar{p}^{\delta+1}-p^{\delta+1}<0$. Hence,
\begin{equation}
\log\frac{H(m,p)}{H(0,p)}\leq -2m[\log(2p)-\varepsilon(m)] + O\left(\frac{1}{M}\right)
\end{equation}
where $\varepsilon(m)=-(2m)^{-1}[\log(n-m+1)-\log(M+1)]$. Since $\log(x)$ is a concave function, we can know that
\begin{equation}
\varepsilon(m)\leq (M)^{-1}\log(M+1)=O\left(M^{-1}\right)
\end{equation}
which concludes the first case in Lemma~\ref{UpBound1}. Similarly, for $2m>M$ (i.e. $n<m$), we can have
\begin{equation}
\log\frac{H(m,p)}{H(M,p)}\leq -2n[\log(2p)-\hat{\varepsilon}(n)] + O\left(\frac{1}{M}\right)
\end{equation}
where $\hat{\varepsilon}(n)\leq O(M^{-1})$. Thereby, we can conclude the second case of Lemma~\ref{UpBound1}. Note that the case where $p<0.5$ can be derived by using Lemma~\ref{Symmetry}.
\end{proof}
\end{lemma}
For the case that $t=0$ and $M\gg 1$, using the same method as the above proof, we can derive the same lower and upper bounds as Lemmas~\ref{UpBound1} and \ref{LowBound1}. On the other hand, for $t=0$, Lemma~\ref{Ratio1} does not hold and we can have
\begin{lemma}
\label{Ratio0}
$H(m,p;M,0)=H(n,p;M,0)$
\begin{proof}
When $t=0$,
\begin{equation}
H(m,p)=2^{M+1}(2p-1)^{-1}\int_{\bar{p}}^{p}x^n(1-x)^m\mathrm{d}x.
\end{equation}
Then, substituting $x$ as $1-v$ concludes Lemma~\ref{Ratio0}.
\end{proof}
\end{lemma}

\section{Proof for Theorem 1}
Following the notations in Section B, when $M\gg 1$ in Eqn. (\ref{equaton:score}), we have $\tilde{\mathbb{P}}_i=\mathbb{E}_{\mathcal{L}}(x_i+z_i)/M+O(1/M)$, where $\mathbb{E}_{\mathcal{L}}$ denotes the expectation of $\mathbb{P}(\mathcal{L}|\bm{L})$. Meanwhile, according to Chebyshev's inequality, $\mathbb{P}_i= (x_i+w_i)/M+\epsilon$, where $|\epsilon|\leq_{1-\delta}O(1/\sqrt{\delta M})$ and $\delta$ is any given number in $(0,1)$.
Here, we use $a\leq_{1-\delta} b$ to denote that $a$ is smaller or equal than $b$ with probability $1-\delta$.
Thus, we can calculate the upper bound of $|\tilde{\mathbb{P}}_i-\mathbb{P}_i|$ as
\begin{equation}
\label{ErrorConnect}
|\tilde{\mathbb{P}}_i-\mathbb{P}_i|\leq_{1-\delta} \mathbb{E}_{\mathcal{L}}|w_i-z_i|/M +O(1/\sqrt{M}) \leq \mathbb{E}_{\mathcal{L}}\left[m/M\right]+O(1/\sqrt{M}).
\end{equation}
Recalling Lemma~1, we know that when $M\gg 1$,
\begin{equation}
\mathbb{E}[m/M]\lesssim (1+e^{\delta})^{-1}(\varepsilon+e^{\delta})(1+\varepsilon)^{M-1}\;,\;
\mathbb{E}[m/M]^2\lesssim (1+e^{\delta})^{-1}(\varepsilon^2+e^{\delta})(1+\varepsilon)^{M-2}.
\end{equation}
where $\varepsilon^{-1}=\prod_{i=0}^{N}(2\hat{\mathbb{P}}_i)^{2}$, $\delta=O[\Delta\cdot \log(M)]$ and $\Delta={\sum}_{i=1}^N[1(\mathbb{P}_i<0.5)-1(\mathbb{P}_i>0.5)]$.
If $\Delta<0$, from the definition of $\Delta$, we can know that $\Delta\leq 1$. Thus, $e^{\delta}\leq O(1/M)$.
Furthermore, when ${\prod}_{i=0}^{N}(2\hat{\mathbb{P}}_{i})^{2} \geq M$, $\varepsilon\leq M^{-1}$.
Thereby, 
\begin{equation}
\label{bounddd}
\mathbb{E}\left[\frac{m}{M}\right]\lesssim \frac{C_{1}}{M\cdot C_2}\;\;, \;\;\mathbb{E}\left[\frac{m}{M}\right]^2\lesssim \frac{C_{1}}{M^2\cdot C_2^2}
\end{equation}
where $C_{1}=(1+M^{-1})^{M}\approx e$ and $C_{2}=1+M^{-1}\approx 1$.
Based on Eqn. (\ref{bounddd}), we can know $\mathbb{E}[m/M]\lesssim O(1/M)$ and $\mathrm{Var}[m/M]\lesssim O(1/M^2)$.
Again, according to Chebyshev's inequality, we can have $ \mathbb{E}_{\mathcal{L}}\left[m/M\right]\leq_{1-\delta}O(1/\sqrt{\delta}M)$, and we can conclude Theorem 1 by taking the upper bound of $\mathbb{E}_{\mathcal{L}}\left[m/M\right]$ into Eqn. (\ref{ErrorConnect}).

\section{Background for Reinforcement Learning}
In this section, we introduce some important concepts about reinforcement learning (RL). In an RL problem, an agent interacts with an unknown environment and attempts to maximize its cumulative collected reward \cite{Sutton98,Szepesvari10}. The environment is commonly formalized as a Markov Decision Process (MDP) defined as $\mathcal{M} = \langle \mathcal{S}, \mathcal{A}, \mathcal{R}, \mathcal{P}, \gamma\rangle$. At time $t$ the agent is in state $s_t \in \mathcal{S}$ where it takes an action $a_t \in \mathcal{A}$ leading to the next state $s_{t+1} \in \mathcal{S}$ according to the transition probability kernel $\mathcal{P}$, which encodes $\mathbb{P}(s_{t+1}\mid s_t,a_t)$. In most RL problems, $\mathcal{P}$ is unknown to the agent. The agent's goal is to learn the optimal policy, a conditional distribution $\pi(a \mid s)$ that  maximizes the sate's value function. The value function calculates the cumulative reward the agent is expected to receive given it would follow the current policy $\pi$ after observing the current state $s_t$
$$
V^\pi(s) = \mathbb{E}_\pi \left[ \sum_{k=1}^\infty \gamma^k r_{t+k} \mid s_t = s   \right].
$$
Intuitively, it measures how preferable each state is given the current policy.

As a critical step towards improving a given policy, it is a standard practice for RL algorithms to learn a state-action value function (i.e. Q-function). Q-function calculates the expected cumulative reward if agent choose $a$ in the current state and follows $\pi$ thereafter
\begin{equation*}
\begin{split}
Q^\pi(s,a) =\mathbb{E}_\pi\left[ \mathcal{R}(s_t,a_t,s_{t+1}) + \gamma V^\pi(s_{t+1}) \mid s_t = s, a_t = a \right].
\end{split}
\end{equation*}
In real-world problems, in order to achieve better generalization, instead of learning a value for each state-action pair, it is more common to learn an approximate value function: $Q^\pi(s,a; \theta) \approx Q^\pi(s,a)$. A standard approach is to learn a feature-based state representation $\phi(s)$ instead of using the raw state $s$ \cite{Gordon00}. Due to the popularity of Deep Reinforcement learning, it has been a trend to deploy neural networks to automatically extract high-level features \cite{Silver17,Mnih15}. However, running most deep RL models are very computationally heavy. On contrast, static feature representations are usually light-weight and simple to deploy. Several studies also reveal that a carefully designed static feature representation can achieve performance as good as the most sophisticated deep RL models, even in the most challenging domains \cite{Liang16}.

\section{Utility-Maximizing Strategy for Workers}
\begin{lemma}
\label{Strategy}
For worker $i$, when $M\gg 1$ and $a_t>\frac{c_{i, H}}{\mathbb{P}_{i,H}-0.5}$, if $\tilde{\mathbb{P}}^t_i\approx \mathbb{P}^t_i$, reporting truthfully ($\textsf{rpt}^{t}_i=1$) and exerting high efforts ($\textsf{eft}^{t}_i=1$) is the utility-maximizing strategy.
\begin{proof}
When $M\gg 1$, we can have$\sum_j \textsf{sc}_i(j)\approx M\cdot \tilde{\mathbb{P}}_i$. Thus, the utility of worker $i$ can be computed as
\begin{equation}
u_i^t\approx  M\cdot a_t\cdot (\tilde{\mathbb{P}}_i-0.5) + M\cdot b- M \cdot c_{i,H}\cdot \textsf{eft}^{t}_i.
\end{equation}
Further considering Eqn. (1) and $\mathbb{P}_L=0.5$, if $\tilde{\mathbb{P}}^t_i\approx \mathbb{P}^t_i$, we can compute worker $i$'s utility as
\begin{equation}
u_i^t\approx M\cdot [a_t(2\cdot \textsf{rpt}^{t}_i-1)(\mathbb{P}_{i,H}-0.5)-c_{i,H}]\cdot\textsf{eft}^{t}_i+ M\cdot b.
\end{equation}
Thereby, if $a_t>\frac{c_{i,H}}{\mathbb{P}_{i,H}-0.5}$, $\textsf{rpt}^{t}_i=1$ and $\textsf{eft}^{t}_i=1$ maximize $u_i^t$, which concludes Lemma~\ref{Strategy}.
\end{proof}
\end{lemma}

\section{Uninformative Equilibrium}
The uninformative equilibrium denotes the case where all workers collude by always reports the same answer to all tasks.
For traditional peer prediction mechanisms, under this equilibrium, all the workers still can get high payments because these mechanisms determines the payment by comparing the reports of two workers.
However, the data requester only can get uninformative labels, and thus this equilibrium is undesired.

In our mechanism, when workers always reports the same answer, for example $1$, our Bayesian inference will 
wrongly regard the collected labels as high-quality ones and calculate the estimates as
\begin{equation}
\tilde{\mathbb{P}}_i=\frac{M+2}{M+3}\;,\;\tilde{\tau}_{-1}=\frac{M+1}{M+2}.
\end{equation}
If the answer is $2$, our estimates are
\begin{equation}
\tilde{\mathbb{P}}_i=\frac{M+2}{M+3}\;,\;\tilde{\tau}_{+1}=\frac{M+1}{M+2}.
\end{equation}
In this case, we can build a warning signal for the uninformative equilibrium as
\begin{equation}
\mathsf{Sig}_u = \frac{1}{N}\sum_{i=1}^{N}\log(\tilde{\mathbb{P}}_i)+\log(\max\{\tilde{\tau}_1,\tilde{\tau}_2 \})
\end{equation}
If
\begin{equation}
\mathsf{Sig}_u\geq \log\frac{M+1}{M+3}
\end{equation}
workers are identified to be under the uniformative equilibrium, and we will directly set the score in our payment rule as $0$.
By doing so, we can create a huge loss for workers and push them to leave this uninformative equilibrium.

\section{Proof for Theorem 3}
In our proof, if we omit the superscript $t$ in an equation, we mean that this equation holds for all time steps.
Due to the one step IC, we know that, to get higher long term payments, worker $i$ must mislead our RIL algorithm into at least increasing the scaling factor from $a$ to any $a'>a$ at a certain state $\hat{s}$.
Actually, our RIL algorithm will only increase the scaling factor when the state-action value function satisfies $Q^{\pi}(\hat{s},a)\leq Q^{\pi}(\hat{s},a')$.
Eqn. (\ref{equation:approx_reward}) tells us that the reward function consists of the utility obtained from the collected labels ($F(\tilde{A}^t)$) and the utility lost in the payment ($\eta {\sum}_{i=1}^{N}P^t_i$).
Once we increase the scaling factor, we at least need to increase the payments for the other $N-1$ workers by $M\sum_{x\neq i} \mathbb{P}_{x,H}\cdot G_{\mathcal{A}}$, corresponding to the left-hand side of the first equation in Eqn. (\ref{Condition}).

On the other hand, for the obtained utility from the collected labels, we have 
\begin{lemma}
\label{ConvBoundxx}
At any time step $t$, if all workers except worker $i$ report truthfully and exert high efforts, we have $F(\tilde{A}^t)\leq F(1)$ and $F(\tilde{A}^t)\geq F(1-\psi)$, where $\psi$ is defined in Eqn. (\ref{Condition}).
\begin{proof}
In our Bayesian inference algorithm, when $M\gg 1$, the estimated accuracy $\tilde{A}$ satisfies
\begin{equation}
\label{accP}
\tilde{A} \approx 1-\mathbb{E}g(\tilde{\sigma}_j)\;\;,\;\; g(\tilde{\sigma}_j)=1/(1+e^{|\tilde{\sigma}_j|}).
\end{equation}
From the proof of Theorem~2, we can know that $\tilde{\mathbb{P}}^{t}_i \approx \mathbb{P}^{t}_i$.
In this case, according to Eqn. (\ref{ProbRatio}), we can have
\begin{equation}
\label{ProbRatioApp}
\tilde{\sigma}_j(\mathbb{P}_i)\approx \log\left(\frac{\tau_{-1}}{\tau_{+1}}\lambda_i^{\delta_{ij1}-\delta_{ij2}}{\prod}_{k\neq i}\lambda_H^{\delta_{kj1}-\delta_{kj2}}\right).
\end{equation}
where $\lambda_i=\mathbb{P}_i/(1-\mathbb{P}_i)$ and $\lambda_H=\mathbb{P}_H/(1-\mathbb{P}_H)$.

We know that $\tilde{A}\leq 1.0$ holds no matter what strategy worker $i$ takes. To prove Lemma~2, we still need to know the lower bound of $\tilde{A}$. Thus, we consider two extreme cases where worker $i$ intentionally provides low-quality labels:

\underline{\textbf{Case 1}:} If $\mathbb{P}_i=0.5$, we can eliminate $\lambda_i$ from Eqn.\ref{ProbRatioApp} because $\lambda_i=1$. Furthermore, according to Lemma~\ref{Concave2}, we can know that 
$g(\tilde{\sigma}_j)< e^{\tilde{\sigma}_j}$ and $g(\tilde{\sigma}_j)< e^{-\tilde{\sigma}_j}$ both hold. Thus, we build a tighter upper bound of $g(\tilde{\sigma}_j)$ by dividing all the combinations of $\delta_{kj1}$ and $\delta_{kj2}$ in Eqn.\ref{ProbRatioApp} into two sets and using the smaller one of $e^{\tilde{\sigma}_j}$ and $e^{-\tilde{\sigma}_j}$ in each set.
By using this method, if the true label is $-1$, we can have $\mathbb{E}_{[L(j)=-1]}g(\tilde{\sigma}_j)< q_1+q_2$, where
\begin{equation*}
\begin{split}
&q_1 = \frac{\tau_{+1}}{\tau_{-1}}{\sum}_{n=K+1}^{N-1}C_{N-1}^{n} (\frac{1}{\lambda_H})^{n-m}\mathbb{P}_H^n(1-\mathbb{P}_H)^m\\
&q_2 = \frac{\tau_{-1}}{\tau_{+1}}{\sum}_{n=0}^{K}C_{N-1}^{n} {\lambda_H}^{n-m}\mathbb{P}_H^n(1-\mathbb{P}_H)^m\\
&n={\sum}_{k\neq i}\delta_{kj(-1)}\;,\;m= {\sum}_{k\neq i}\delta_{kj(+1)}
\end{split}
\end{equation*}
and $K=\lfloor (N-1)/2 \rfloor$. By using Lemma~\ref{Concave3}, we can thus get
\begin{equation*}
\begin{split}
\mathbb{E}_{[L(j)=-1]}g(\tilde{\sigma}_j) < c_{\tau}[4\mathbb{P}_H(1-\mathbb{P}_H)]^{\frac{N-1}{2}}.
\end{split}
\end{equation*}
where $c_{\tau}=\tau_{-1}\tau_{+1}^{-1}+\tau_{-1}^{-1}\tau_{+1}$. Similarly,
\begin{equation*}
\begin{split}
\mathbb{E}_{[L(j)=+1]}g(\tilde{\sigma}_j) < c_{\tau}[4\mathbb{P}_H(1-\mathbb{P}_H)]^{\frac{N-1}{2}}.
\end{split}
\end{equation*}
Thereby, $\tilde{A}>1-c_{\tau}[4\mathbb{P}_H(1-\mathbb{P}_H)]^{\frac{N-1}{2}}=1-\psi$.

\underline{\textbf{Case 2}:} If $\mathbb{P}_i=1-\mathbb{P}_H$, we can rewrite Eqn.\ref{ProbRatioApp} as
\begin{equation*}
\tilde{\sigma}_j(1-\mathbb{P}_H)\approx \log\left(\frac{\tau_{-1}}{\tau_{+1}}\lambda_H^{x-y}{\prod}_{k\neq i}\lambda_H^{\delta_{kj(-1)}-\delta_{kj(+1)}}\right)
\end{equation*}
where $x=\delta_{ij(+1)}$ and $y=\delta_{ij(-1)}$. Since $\mathbb{P}_i=1-\mathbb{P}_H$, $x$ and $y$ actually has the same distribution as $\delta_{kj(-1)}$ and $\delta_{kj(+1)}$. Thus, the distribution of $\tilde{\sigma}_j(1-\mathbb{P}_H)$ is actually the same as $\tilde{\sigma}_j(\mathbb{P}_H)$.
In other words, since Theorem~2 ensures $\tilde{\mathbb{P}}_i\approx\mathbb{P}_i$, our Bayesian inference algorithm uses the information provided by worker $i$ via flipping the label when $\mathbb{P}_i<0.5$.

Thus, even if worker $i$ intentionally lowers the label quality, $\tilde{A}\geq 1-\psi$ still holds.
Considering $F(\cdot)$ is a non-decreasing monotonic function, we conclude Lemma~2.
\end{proof}
\end{lemma}
Thereby, if Eqn. (13) is satisfied, worker $i$ will not be able to cover $Q$ value loss in the payments, and our RL algorithm will reject the hypothesis to increase the scaling factor.
In this case, the only utility-maximizing strategy for worker $i$ is to report truthfully and exert high efforts.

\section{Worker Models}
To demonstrate the general applicability of our mechanism, we test it under three different worker models in Section 5.2, with each capturing a different way to decide the labeling strategy.
The formal description of the three models is as follows:
\begin{itemize}[topsep=0pt, partopsep=0pt]
\item {\bf Rational} workers alway act to maximize their own utilities. Since our incentive mechanism theoretically ensures that exerting high effort 
is the utility-maximizing strategy for all workers (proved in Section~4), it is safe to assume workers always do so as long as the payment is high enough to cover the cost.
\item {\bf Quantal Response (QR)} workers \cite{mckelvey1995quantal} exert high efforts with the probability 
$$
\textsf{eft}_i^t= \frac{\exp(\lambda\cdot  u_{iH}^t)}{\exp(\lambda \cdot u_{iH}^t) + \exp (\lambda \cdot u_{iL}^t)}
$$
where $u_{iH}^t$ and $u_{iL}^t$ denote worker $i$'s expected utility after exerting high or low efforts respectively at time $t$. $\lambda$ describe workers' rationality level and we set $\lambda =3$.

\item {\bf Multiplicative Weight Update (MWU)} workers \cite{chastain2014algorithms} update their probabilities of exerting high efforts at every time step $t$ after receiving the payment as the following equation
\begin{align*}
\textsf{eft}_i^{t+1} = \frac{\textsf{eft}_i^t(1+\bar{u}_{\cdot H})}{\textsf{eft}_i^t(\bar{u}_{\cdot H} - \bar{u}_{\cdot L}) + \bar{u}_{\cdot L} + 1}
\end{align*}
where $\bar{u}_{\cdot H}$ and $\bar{u}_{\cdot L}$ denote the average utilities received if exerting high efforts or low efforts at time $t$ respectively. We initialize $\textsf{eft}_i^0$ as $0.2$ in our experiments.
\end{itemize}
\end{document}